\documentclass[aps,pra,reprint,superscriptaddress,amsmath,amssymb,showkeys]{revtex4-1}

\usepackage{xcolor}

\providecommand{\abs}[1]{\lvert#1\rvert}
\DeclareMathOperator{\tr}{tr}

\usepackage{tabularx}
\usepackage{xspace}
\newcommand{\lisa}{LISA\xspace}
\newcommand{\drops}{DROPS\xspace}

\usepackage{graphicx} % Required for including pictures
\usepackage{float}
\usepackage{bbold}
\usepackage{bm}
\usepackage{amsthm}
\newtheorem{theorem}{Theorem}
\newtheorem{result}[theorem]{Result}
\usepackage[mathscr]{eucal}

\usepackage[bookmarks=false,pdfstartview={FitH}]{hyperref}

\begin{document}

%-----------------------------------------------------------------------------------------------------
%---- general informations
\title{Wigner tomography of multispin quantum states}
\author{David Leiner}
\email[]{david.leiner@tum.de}
\author{Robert Zeier}
\email[]{zeier@tum.de}
\author{Steffen J. Glaser}
\email[]{steffen.glaser@tum.de}
%\thanks{}
\affiliation{Technische Universit\"at M\"unchen, Department Chemie, Lichtenbergstrasse 4, 
85747 Garching, Germany}
%\date{\today}
\date{January 15, 2018}
%-----------------------------------------------------------------------------------------------------

%-----------------------------------------------------------------------------------------------------
%---- Abstract
\begin{abstract}
We study the tomography of multispin quantum states
in the context of finite-dimensional Wigner representations.
An arbitrary operator can be completely characterized and visualized
using multiple shapes assembled from linear combinations of spherical harmonics
[A. Garon, R. Zeier, and S. J. Glaser, Phys.\ Rev.\ A 91, 042122 (2015)].
We develop a general methodology to experimentally recover 
these shapes by measuring expectation values of rotated 
axial spherical tensor operators and provide
an interpretation in terms of fictitious multipole potentials.
Our approach is experimentally demonstrated 
for quantum systems consisting of up to three spins 
using nuclear magnetic resonance spectroscopy.  
\end{abstract}
%-----------------------------------------------------------------------------------------------------

%-----------------------------------------------------------------------------------------------------
%---- PACS number, Keywords
%\pacs{}
\maketitle
%----------------------------------------------------------------------------------------------------

%-----------------------------------------------------------------------------------------------------
%----Introcution
%-----------------------------------------------------------------------------------------------------
\section{Introduction}

Optical homodyne tomography can be applied to experimentally measure
the quantum state of light \cite{smithey1,smithey2,smithey3,leonhardt,paris_rehacek}.
One thereby recovers 
an infinite-dimensional Wigner function \cite{wigner1932, SchleichBook, Curtright-review}
as a classically motivated phase-space representation, 
providing a useful tool for the characterization and visualization of 
quantum-optical systems \cite{Nussenzveig}.
This results in an advantageous dualism between measurement scheme
and phase-space representation, which we would like to transfer to the case of 
finite-dimensional, \emph{coupled} spin systems.

One important representation of finite-dimensional quantum systems 
relies on discrete Wigner functions 
\cite{Wooters87,leonhardt1996,Miquel,Miquel2,gibbons2004,FE09}.
But we will restrict ourselves to continuous representations
in order to naturally reflect the inherent rotational symmetries
of spins. Individual spins are faithfully described by their
magnetization vector (or Bloch vector), which, however, neglects
relevant parts of the full density matrix in the case of multiple, coupled spins.
These missing parts include zero- and multiple-quantum or antiphase
coherence  as well as spin alignment \cite{EBW87},
which are partially characterized by visual approaches based 
on single-transition operators \cite{EBW87,Donne_Gorenstein,Freeman97}.

We will follow the general strategy of Stratonovich \cite{Stratonovich}
which specifies criteria for the definition of 
continuous Wigner functions for finite-dimensional quantum systems. 
The case of single spins is widely studied in the literature
\cite{VGB89,Brif98,StarProd,klimov2005classical,Brif97,klimov2002ExactEvolution},
and visualizations for multiple spins have been considered 
in \cite{DowlingAgarwalSchleich,SchleichBook,JHKS,PhilpKuchel,Harland} with
various degrees of generality. However, until very recently, it was not
clear \cite{PhilpKuchel,Harland} if a general Wigner representation also exists
for arbitrary, coupled spin systems, and even the case
of three coupled spins $1/2$ was open.

Fortunately, a general Wigner representation for 
characterizing and visualizing arbitrary coupled spin systems has been developed
in \cite{Garon15}. It is based on mapping arbitrary operators to a set of spherical functions
which are denoted as droplets, while preserving crucial features of the quantum system.
The characteristic shapes of these droplets can be interpreted as the result
of an abstract mapping, but we also ask in this paper how they are related to experimentally measurable
quantities. The general Wigner representation introduced
in \cite{Garon15} is denoted as
the \drops representation (discrete representation of
operators for spin systems), 
and its basics are recalled in Sec.~\ref{drops_sum} where important properties
are also summarized.

In this paper, we theoretically develop 
a tomography scheme for spherical functions of arbitrary multispin quantum states.
We study experimental schemes to reconstruct 
the generalized Wigner representation of a given density operator 
(representing mixed or pure quantum states). 
Extensions to quantum process tomography \cite{PhysRevA.64.012314}
as given by the experimental reconstruction of entire 
propagators (e.g., representing quantum gates) are, however, beyond the scope of this paper.
Our scheme is particularly tailored to
the Wigner representation of \cite{Garon15}, for which 
an interpretation in terms of
fictitious multipole potentials is provided.
We will focus on systems consisting 
of spins $1/2$, even though our approach is applicable to arbitrary spin numbers. 
We also provide explicit experimental protocols for our Wigner tomography scheme
and demonstrate its feasibility using 
nuclear magnetic resonance (NMR) experiments.
Motivated by our experiments, most  of the discussed examples consider only the traceless part
of the density matrix. 

This paper is organized as follows.  A brief summary of the 
DROPS representation is presented in Sec.~\ref{drops_sum}.
Our general methodology for sampling spherical functions
of multispin operators is introduced in Sec.~\ref{drops_tomo}, which also
states the main technical results for the Wigner tomography.
Section~\ref{interpretation} provides a physical interpretation
of spherical functions 
in terms of fictitious multipole potentials. The performed NMR experiments
are summarized in Sec.~\ref{exp_res}, and Sec.~\ref{temporal}
discusses the use of temporal averaging.
The precise experimental scheme and its implementation on a spectrometer
are detailed in Secs.~\ref{theory} and \ref{exp_impl}.
We conclude by 
summarizing and discussing theoretical 
and experimental 
aspects, while also contrasting our paper with other 
tomography approaches.
Further details are deferred to the Appendices.

%-----------------------------------------------------------------------------------------------------
%---- Theory
%-----------------------------------------------------------------------------------------------------

\section{Visualization of operators using spherical functions\label{drops_sum}}
\begin{figure}
\includegraphics{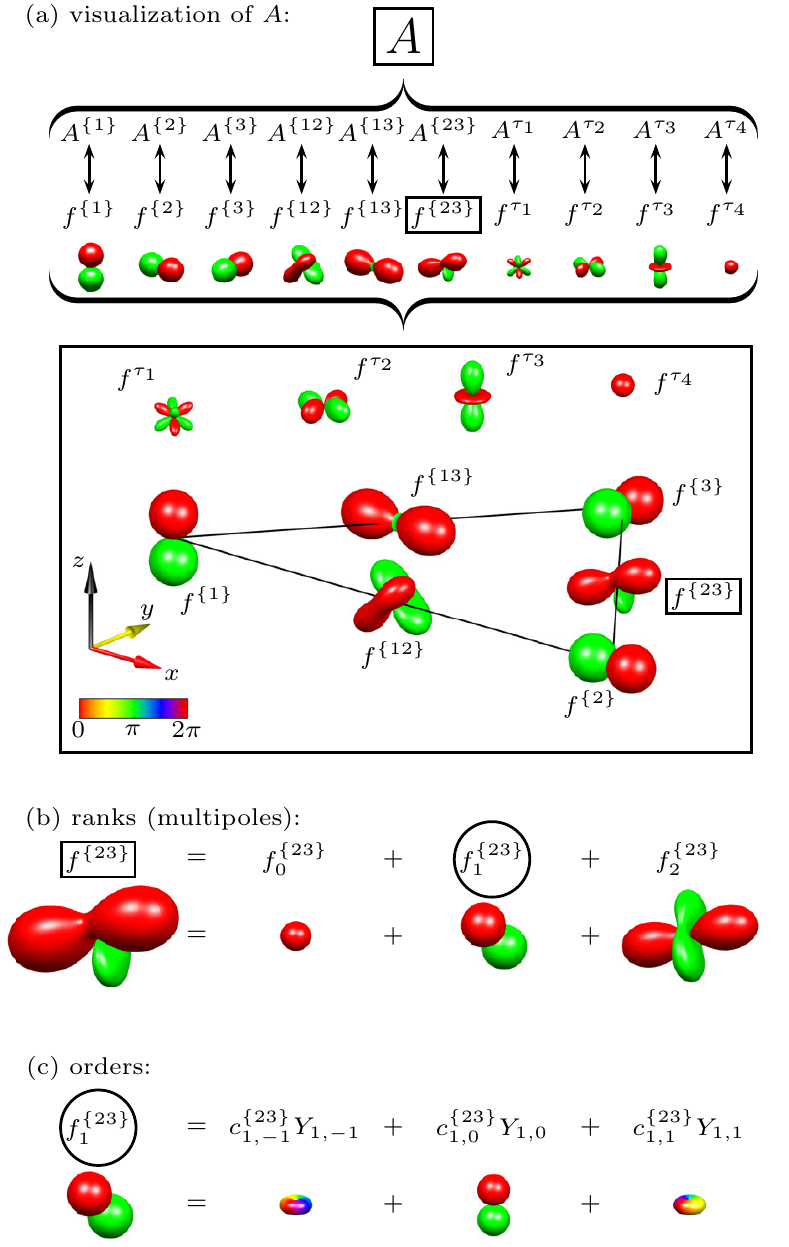}
\caption{(Color online) (a) Three-spin operator
$A=I_{1z}+I_{2x}+I_{3y}+2I_{1x}I_{2z}+I_{2x}I_{3x}+I_{2x}I_{3y} +
I_{2x}I_{3z}+2 I_{1x}I_{3x}+4I_{1x}I_{2x}I_{3x}$
\cite{EBW87,Note\thecart}
visualized using multiple spherical functions $f^{(\ell)}=f^{(\ell)}(\theta,\phi)$,
and individual components $A^{(\ell)}$ of $A$ mapped 
to  $f^{(\ell)}$ and graphically visualized;
$A^{\emptyset}$ is here zero;
trilinear labels ``$\{123\},\tau_p$" are shortened to ``$\tau_p$".
(b)~$f^{\{23\}}$ (box)
decomposed into its
$2^j$-multipole contributions $f^{\{23\}}_j$ 
with $j \in \{0,1,2\}$ (monopole, dipole, and quadrupole)
(c)~$f^{\{23\}}_1$ (circle) decomposed
into spherical harmonics of order $m\in\{-1,0,1\}$;  $Y_{1,-1}$ and $Y_{1,1}$ 
are rainbow colored \cite{Note\thecolors}.
\label{fig:mapping_new} } 
\end{figure}

We summarize the approach of \cite{Garon15}
to obtain a Wigner representation of arbitrary operators $A$ 
in coupled spin systems 
using multiple spherical functions, which is
based on a general one-to-one mapping from
spherical tensor operators to spherical harmonics.
An operator
\begin{equation}
\label{A}
A = \sum_{\ell \in L} A^{(\ell)}
\end{equation}
is decomposed according to a suitable set $L$ of labels $\ell$ (i.e.\ quantum numbers)
inducing a bijective mapping
between the components $A^{(\ell)}$ 
and spherical functions $f^{(\ell)}=f^{(\ell)}(\theta,\phi)$.
These spherical functions
can be plotted together as seen in the example of Fig.~\ref{fig:mapping_new}(a) where
the corresponding mapping is highlighted.
This provides a pictorial representation of the operator $A$, which
conserves important properties  and symmetries 
depending on the chosen label set $L$.

\addtocounter{footnote}{1}
\footnotetext[\value{footnote}]{
The components  $T_{jm}$ of 
varying rank $j\in\{0,1,\ldots,2s\}$ and 
order $m\in\{-j,\ldots,j\}$ form a complete orthonormal 
matrix basis.  In particular, one obtains the matrix basis
$
T_{00} = 
\left(
\begin{smallmatrix}
1 & 0\\
0 & 1
\end{smallmatrix}\right)/\sqrt{2}$,
$T_{1,-1} = 
\left(
\begin{smallmatrix}
0 & 0\\
1 & 0
\end{smallmatrix}
\right)$,
$T_{10} = 
\left(
\begin{smallmatrix}
1 & 0\\
0 & -1
\end{smallmatrix}
\right)/\sqrt{2}$,
$T_{11} = 
\left(
\begin{smallmatrix}
0 & -1\\
0 & 0
\end{smallmatrix}
\right)
$
for a single spin with spin number $s=1/2$.}
\newcounter{tensor}
\setcounter{tensor}{\value{footnote}}

\addtocounter{footnote}{1}
\footnotetext[\value{footnote}]{Spherical harmonics
$Y_{jm}(\theta,\phi)=r(\theta,\phi) \exp[i\eta(\theta,\phi)]$
(and spherical functions) are plotted 
throughout this work by mapping their spherical coordinates $\theta$ and $\phi$ to the radial 
part $r(\theta,\phi)$ and phase $\eta(\theta,\phi)$.}
\newcounter{spherical}
\setcounter{spherical}{\value{footnote}}

The components $A^{(\ell)}$ 
and the spherical functions $f^{(\ell)}=f^{(\ell)}(\theta,\phi)$
can be further split up into their multipole contributions  $A_{j}^{(\ell)}$ and 
$f_{j}^{(\ell)}=f^{(\ell)}_j(\theta,\phi)$ depending on the ranks $j\in J(\ell)$
occuring for each label $\ell$ as shown in Fig.~\ref{fig:mapping_new}(b), i.e.,
\begin{equation}\label{Aj}
A^{(\ell)}  = \sum_{j \in J(\ell)} A_{j}^{(\ell)} \;\text{ and }\;
f^{(\ell)}  = \sum_{j \in J(\ell)} f_{j}^{(\ell)}.
\end{equation}
Finally, the rank-$j$ multipole contributions [see Fig.~\ref{fig:mapping_new}(c)]
\begin{equation}\label{Aj_fj}
A^{(\ell)}_j  = \sum_{m=-j}^{j} c_{jm}^{(\ell)} T_{jm}^{(\ell)}
\;\text{ and }\;
f^{(\ell)}_j  = \sum_{m=-j}^{j} c_{jm}^{(\ell)} Y_{jm},
\end{equation}
can be decomposed into
components of irreducible spherical
tensor operators $T_{jm}^{(\ell)}$ \cite{Wigner31,Wigner59,Racah42,BL81,Silver76,CH98}
 and the corresponding spherical harmonics  $Y_{jm}$ \cite{Jac99,Note\thespherical}
of order $ m$ with $-j\leq m\leq j$. Note the identical
expansion coefficients $c_{jm}^{(\ell)}$ in Eq.~\eqref{Aj_fj}.
The dualism in Eqs.~\eqref{Aj} and \eqref{Aj_fj} exploits the well-known correspondence between 
irreducible tensor operators and spherical harmonics \cite{Silver76,CH98}.
In summary, an operator $A$ is mapped to a set of spherical functions
$f^{(\ell)}$, each of which is referred to as a droplet identified by $\ell$.
The whole representation (and its visualization)
was introduced in \cite{Garon15} and is denoted as
the \drops representation (discrete representation of
operators for spin systems), and it lends 
itself to interactively exploring the dynamics of multispin systems, e.g., by 
use of the free application \cite{ipad_app}.

\addtocounter{footnote}{1}
\footnotetext[\value{footnote}]{
Cartesian operators for single spins are 
$I_x:=\sigma_x/2$, $I_y:=\sigma_y/2$, and $I_z:=\sigma_z/2$, where
the Pauli matrices are
$
\sigma_x=\left(
\begin{smallmatrix}
0 & 1\\
1 & 0
\end{smallmatrix}
\right)
$,
$
\sigma_y=\left(
\begin{smallmatrix}
0 & -i\\
i & 0
\end{smallmatrix}
\right)
$, and
$
\sigma_z=\left(
\begin{smallmatrix}
1 & 0\\
0 & -1
\end{smallmatrix}
\right)
$. For $n$ spins, one has the operators
$I_{k \eta} := \bigotimes_{s=1}^{n} I_{a_{s}}$ 
where $a_{s}$ is equal to $\eta$ for $s{=}k$ 
and is zero otherwise; note $I_{0}:=
\left(\begin{smallmatrix}
1 & 0\\
0 & 1
\end{smallmatrix}\right)$.}
\newcounter{cart}
\setcounter{cart}{\value{footnote}}

\addtocounter{footnote}{1}
\footnotetext[\value{footnote}]{Hermitian operators
lead to positive and negative values which 
are shown in red (dark gray) and green (light gray).}
\newcounter{colors}
\setcounter{colors}{\value{footnote}}

The example presented in Fig.~\ref{fig:mapping_new} 
uses one particular version of this representation 
which relies on the \lisa tensor operator basis
as defined in \cite{Garon15}, which is characterized by the 
\underline{li}nearity of the basis operators, the involved {\underline s}ubsystem, 
and {\underline a}uxiliary criteria, such as permutation symmetry. 
For coupled spins $1/2$, operators are first decomposed in this basis
according to the set of involved spins, e.g.,
one introduces the labels $\{k\}$ and $\{kl\}$
for linear and bilinear operators 
acting on a subset of one or two spins numbered by $k$ and $l$, and so forth.
Secondly, the LISA basis for operators acting on three or more spins
needs to also distinguish symmetry properties under permutations, i.e., 
combined labels such as ``$\{klm\},\tau_p$" are used, where the permutation 
symmetry type $\tau_p$ is given by a Young tableau~\cite{Sagan01}. Finally, 
further \textit{ad hoc} labels are necessary for operators involving six or more spins.
Arbitrary operators of a coupled spin system can be uniquely represented using
this \lisa tensor operator basis. Additional details for the visualization technique 
are given in \cite{Garon15} which also discusses
alternative labeling approaches for \drops representations.

The presented Wigner representation can be applied to general mixed 
quantum states as represented by the density operator, and it 
is not limited to pure quantum states as given by 
a state function. In fact, it can be used to 
represent arbitrary operators of spin systems: examples include Hermitian operators 
as Hamiltonians or
density operators representing observables as well as
non-Hermitian operators such as propagators or general quantum gates
\cite{Garon15}.

The
Wigner representation using the \lisa basis is particularly attractive for the 
visualization and analysis of quantum states in magnetic resonance spectroscopy 
\cite{EBW87} and quantum information processing \cite{NC00} and its properties 
have been discussed in \cite{Garon15}.\\
(a)~The \emph{location} of the droplets 
informs about which and how many spin operators
are involved in a given quantum state and what symmetries 
under particle exchange are present [c.f.\ Fig.~\ref{fig:mapping_new}(a)].\\
(b)~The \emph{shape} and \emph{colors} of the droplets reflect 
spectroscopically important properties. For example,  states with defined 
coherence order $p$ \cite{EBW87} can be recognized by their axially symmetric 
shape, and the magnitude as well as sign of the coherence order $p$ 
are represented by the number and direction of ``rainbows'' per  revolution around the 
$z$ axis [see Figs.~\ref{over_b} and \ref{over_b2}]. This also allows us to 
recognize characteristic states, such as inphase and anti-phase 
coherences. \\
(c)~Furthermore, our representation directly 
depicts information about reduced density matrices and thereby conveys information related to entanglement 
measures, which would have to be first computed from the density-matrix description
via partial traces. In particular, the size of the droplets corresponding to linear 
terms [positioned in the vertices of the triangle in Fig.~\ref{fig:mapping_new}(a)] 
provides information on the amount of bipartite entanglement measured by 
the concurrence (see Sec.~IV E in \cite{Garon15}). 
This is an example of the fact that relevant information is often determined
already by a subset of all droplets in the \lisa basis. The \lisa basis 
thereby offers a more structured picture than the density matrix, 
even though the number of droplets grows rapidly with increasing
number of spins. However, as pointed out in \cite{Garon15}, this number
grows less rapidly than the number of density matrix elements.\\
(d)~The droplets \emph{rotate}  under 
non-selective pulses in a natural way. In combination with the 
characteristic droplet shapes, this property makes it in many cases possible  
to design experiments that transfer  a given initial state into a desired 
target state without detailed calculations. Beyond merely interpreting 
the occurring characteristic shapes as a result of an abstract mapping, it 
is interesting to ask whether they are connected to experimentally 
measurable quantities.

\section{Sampling spherical functions of multi-spin operators\label{drops_tomo}}
We explain now how the shape of spherical functions 
can be characterized by
suitable chosen spherical samples.
This will be particularly relevant for spherical functions representing
spin operators as discussed in Sec.~\ref{drops_sum}
for which these spherical samples
can be experimentally measured.
One obtains
a reconstruction method for the quantum state in terms of spherical functions.
In the general case, the associated rank-$j$ components $g_j(\theta,\phi)$ 
of an arbitrary spherical function $g(\theta,\phi)$ are determined by
its scalar product with 
rotated versions $R_{\alpha\beta} Y_{j0}(\theta,\phi)$
of axial spherical harmonics $Y_{j0}(\theta,\phi)$, which have 
rank $j$ and order zero. Given two spherical functions $h(\theta,\phi)$ and $g(\theta,\phi)$, 
we recall the definition of their scalar product 
$\langle h(\theta,\phi)\vert g(\theta,\phi)\rangle_{L^2} := \int_{\theta=0}^\pi \int_{\phi=0}^{2 \pi} 
h^\ast(\theta, \phi)\,  g(\theta, \phi) \sin \theta\,   {\rm d} \theta\, {\rm d} \phi$.
The rotation operator $R_{\alpha\beta}$
acts on a spherical function $h(\theta,\phi)$
by first rotating it around the $y$ axis by a polar angle $\beta$
and then rotating the result around the $z$ axis by an azimuthal angle $\alpha$, i.e., 
$R_{\alpha \beta} h(\theta, \phi):=
h[R^{-1}_{\alpha \beta}(\theta, \phi)]=h(\theta {-} \beta, \phi{-}\alpha)$. 
After these preparations, the mathematical result underpinning our reconstruction method
states that the value of the rank-$j$ component $g_j(\beta, \alpha)$ 
is proportional to the scalar product of ${R}_{\alpha\beta} Y_{j0}(\theta, \phi)$ with $g(\theta, \phi)$:

\begin{result}
\label{result1}
Consider a spherical function $g(\theta,\phi)=\sum_{j} g_{j}(\theta,\phi)$.
The rank-$j$ components
$g_{j}(\beta,\alpha)$ for angles $\beta$ and $\alpha$
can be obtained
from the scalar products
\begin{equation}
\label{tomoprimeaa}
g_j(\beta, \alpha)   =   s_j\; 
\langle R_{\alpha\beta} Y_{j0}(\theta,\phi) | g(\theta,\phi)\rangle_{L^2}
\end{equation}
with $s_j:=\sqrt{(2j{+}1)/(4\pi)}$.
\end{result}

Assuming that an operator $A$ is represented by a set of spherical functions 
$f^{(\ell)}(\theta,\phi)$, we can apply Result~\ref{result1} by setting 
$g(\theta,\phi):= f^{(\ell)}(\theta,\phi)$ for each label $\ell$ separately. 
We extend Result~\ref{result1} such that the spherical rank-$j$ components 
$f^{(\ell)}_j(\beta, \alpha)$ can also be recovered by comparing the operator 
$A$ directly with rotated axial tensor operators $\mathcal{R}_{\alpha\beta} T_{j0}^{(\ell)}$. 
Consequently, the values of the rank-$j$ spherical components $f^{(\ell)}_j(\beta, \alpha)$
can be experimentally measured for any combination
of polar angles  $\beta$ 
and azimuthal angles $\alpha$.
Here, $\mathcal{R}_{\alpha\beta} C := \mathfrak{R}_{\alpha\beta} C { \mathfrak{R}}^{-1}_{\alpha\beta}$
describes the rotation of an $n$-spin operator $C$
where the simultaneous rotation $\mathfrak{R}_{\alpha\beta}=
e^{- i \alpha F_z} e^{- i \beta F_y}$ of all spins is defined using the  
total spin operators  $F_{z}= \sum_{k=1}^n I_{kz}$ and 
$F_{y}= \sum_{k=1}^n I_{ky}$ \cite{EBW87,Note\thecart}.   We recall the scalar product 
$\langle C\vert B\rangle= \tr( C^\dagger B)$ for operators 
$C$ and $B$ as well as the definition of the expectation value 
$\langle B \rangle_{\rho}=\tr(\rho B)$
of an operator $B$ if the state of the spin system is given by
the density matrix $\rho$. Our result for
recovering rank-$j$ droplet components of an operator can now be stated as follows:

\begin{result}
\label{result2}
Consider a multi-spin operator $A$ which is represented by a set of
spherical functions $f^{(\ell)}(\theta,\phi)=\sum_{j \in J(\ell)} f_{j}^{(\ell)}(\theta,\phi)$.
For each label $\ell$, the rank-$j$ component $f_{j}^{(\ell)}(\beta,\alpha)$
can be experimentally measured 
for arbitrary angles $\beta$ and $\alpha$ via the scalar products
\begin{equation}
\label{tomoprime}
f_j^{(\ell)}(\beta, \alpha)   =   s_j\;
\langle \mathcal{R}_{\alpha\beta} T_{j0}^{(\ell)} | A\rangle.
\end{equation}
If the density matrix $\rho$ of a spin system can be prepared to be identical to the operator $A$,
the rank-$j$ droplet components are given by the expectation values
\begin{equation}
\label{exptomoprime}
f^{(\ell)}_j(\beta, \alpha)=  
s_j\;   \langle  \mathcal{R}_{\alpha\beta} T^{(\ell)}_{j0} \rangle_\rho.
\end{equation}
\end{result}

The proofs of Results~\ref{result1} and \ref{result2} are deferred to 
Appendices~\ref{sec:proof_sf} and \ref{proof_tens}.
Equation~\eqref{exptomoprime} implies that the rank-$j$ droplet components $f^{(\ell)}_j(\beta, \alpha)$ 
for a density matrix $\rho$ can be calculated from the expectation values of rotated
axial tensor operators $\mathcal{R}_{\alpha\beta} T^{(\ell)}_{j0}$. 
Result~\ref{result2} shows that
one can retrace the shapes of the spherical functions $f^{(\ell)}(\beta, \alpha)$
representing any operator that can be mapped onto the density matrix
if one experimentally measures $f^{(\ell)}(\beta, \alpha)$ for sufficiently many angles $\beta$ and $\alpha$.

\section{Droplets as multipole potentials\label{interpretation}}
The methodology of Wigner tomography as presented in Sec.~\ref{drops_tomo}
can be motivated by relating spherical functions to physical multipole potentials.
Section~\ref{subsec_interpret_dipol} details connections to dipole potentials,
which is then generalized to fictitious multipole potentials in Sec.~\ref{subsec_interpret_fict}.
This allows us to interpret the proposed Wigner tomography
as measuring a fictitious potential using axial multipole sensors
(see  Sec.~\ref{subsec_interpret_sampl}).

\subsection{Spherical functions and dipole potentials\label{subsec_interpret_dipol}}

\begin{figure}[b]
\includegraphics{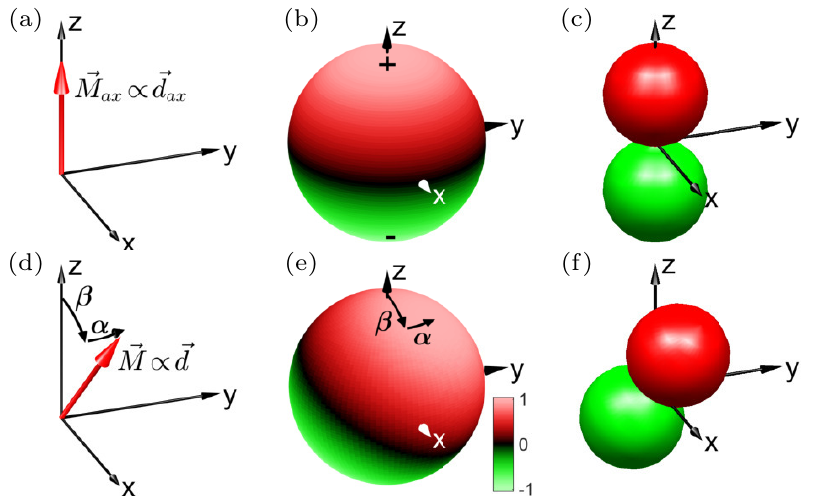}
\caption{(Color online) (a) Axial magnetization vector $\vec{M}_{ax}$ 
and collinear axial dipole vector $\vec{d}_{ax}$. 
(b) Corresponding dipole potential visualized on a sphere, 
with magnitude and sign specified by
brightness and color.
(c) Similar to panel (b), but with magnitude specified by the distance from the origin. (d-f) 
Rotated nonaxial dipole $\vec{d}$ \cite{Note\thecolors}.
\label{fig_pot}}
\end{figure}

The most direct physical interpretation of spherical functions is found for Hermitian 
single-spin terms \cite{EBW87,Note\thecart}
\begin{equation}
\label{rhok}
\rho_k =m_{x} I_{kx}+m_{y} I_{ky}+m_{z} I_{kz}
\end{equation}
of the density matrix
with (possibly time-dependent) real coefficients $m_{x}$, $m_{y}$,
and $m_{z}$.
The corresponding spherical function $f^{\{k\}}(\theta, \phi)$
is now related to a magnetic 
dipole potential.
The operator $\rho_k$ associated with spin $k$ is interpreted as a 
magnetization vector  (or Bloch vector) ${\vec M}= (m_{x} , m_{y} , m_{z} )^T$ 
the components of which are 
proportional to the expectation values of the spin operators $I_{kx}$, 
$I_{ky}$, and $I_{kz}$.
An actual (time-dependent) magnetic dipole  ${\vec d} \propto {\vec M}$
creates a detectable signal  in an NMR experiment
by inducing a voltage in a detection coil.
It is associated  with a scalar dipole potential $V_1(\vec r)$
at $\vec r=\abs{\vec{r}} (\sin\theta \cos \phi, \hskip 0.1em\sin\theta \sin \phi,\hskip 0.1em \cos\theta)$, where 
$\theta$ and $\phi$ are polar and azimuthal angles, respectively.
At a constant distance $\abs{\vec{r}}$ 
from the dipole, the potential $V_1(\vec{r})$ is proportional to the scalar 
product ${\vec d} \cdot {\vec r}$ \cite{Jac99}. In the case of an axial
dipole ${\vec d}_{ax} \propto (0,0,1)^T$ oriented along the $z$ axis, 
the dipole potential is proportional to the axial spherical harmonic
$Y_{10}(\theta, \phi)=\sqrt{3/(4 \pi)}\cos \theta$ as detailed
in Fig.~\ref{fig_pot}(a)-(c).
For a general dipole
${\vec d}={R}_{\alpha \beta}\, {\vec d}_{ax}
\propto
(\sin\beta \cos \alpha,\, \sin\beta \sin \alpha,\, \cos\beta)^T$,
the dipole potential $V_1(\vec{r}) \propto {R}_{\alpha \beta} 
Y_{10}(\theta, \phi)$ is rotated accordingly as shown in 
Fig.~\ref{fig_pot}(d)-(f). Recall that  ${R}_{\alpha \beta}$ denotes a rotation 
around the $y$ axis 
by a polar angle $\beta$
followed by one around the $z$ axis by an azimuthal angle $\alpha$.

\begin{figure}[tb]
\includegraphics{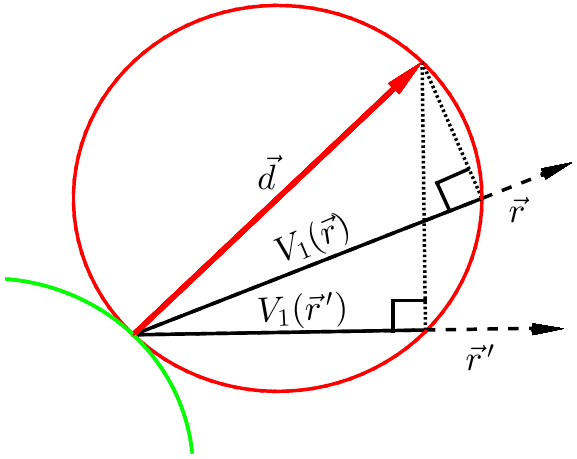}
\caption{(Color online) Slice of Fig.~\ref{fig_pot}(f):
dipole potential $V_1({\vec r})$ 
proportional 
to scalar product   
${\vec d} \cdot {\vec r}$ (projection of ${\vec d}$ onto ${\vec r}$)
due to Thales' theorem, and
similar for $\vec{r}^{\,\prime}$,
results are shown in positive (red) and negative (green) 
spheres in Fig.~\ref{fig_pot}(f); for $\measuredangle\{{\vec d},
{\vec r}\} > 90$ deg, the negative scalar product leads
to the negative sphere.
\label{thales}}
\end{figure}

A scalar dipole potential $V_1(\vec{r})=V_1 (\theta, \phi)$ can be indicated by its 
values on the surface of a  sphere by encoding its sign by the color
and its magnitude by the brightness [see Fig.~\ref{fig_pot}(b) and (e)].
Alternatively, its magnitude can be represented 
by the distance from the origin as in Fig.~\ref{fig_pot}(c) and (f),
where dipole potentials are shown as 
a positive, red (dark gray) sphere and a negative, green (light gray) one
which touch each other at the origin.
This characteristic shape arises as
$V_1 (\theta, \phi)$ is proportional   
to the projection of the dipole ${\vec d}$ onto ${\vec r}$ as depicted in  Fig.~\ref{thales}.
Also, the vector from the center of the negative sphere to the positive one
is collinear with ${\vec d}$.

In summary, a single-spin axial spherical tensor operator $T_{10}^{\{k\}} = \sqrt{2} I_{kz}$
is mapped to the 
axial spherical harmonics $Y_{10}(\theta, \phi)$, and $\rho_k$ from  Eq.~\eqref{rhok}
is mapped to
\begin{equation}
\label{RotDrops}
f^{\{k\}}(\theta, \phi) = \abs{\vec{M}} \, 
{R}_{\alpha \beta} Y_{10}(\theta, \phi) / \sqrt{2},
\end{equation}
where $\beta= {\rm atan}(m_{z}/\sqrt{m^2_{x}+m^2_{y}} )$, $\alpha= {\rm atan} (m_{y}/m_{x})$, and
$\abs{\vec{M}}=\sqrt{m^2_x+m^2_y+m^2_z}$.
Although the direct correspondence between spherical functions
and actual physical dipole potentials appears to be limited to the case
of single-spin terms, it suggest the following interpretation for other
spherical functions presented here.

\subsection{Fictitious multipole potentials\label{subsec_interpret_fict}}
Any spherical function $f^{(\ell)}(\theta, \phi)$ can be regarded as the 
potential $V^{(\ell)}(\vec{r})$ of a fictitious charge distribution $\sigma^{(\ell)} (\vec{r})$
localized in a small volume close to the origin, i.e., $\sigma^{(\ell)}(\vec{r})$ is 
non-zero only  for $| \vec{r}| \ll 1$. At a radius of $| \vec{r}| = 1$,
the potential  can be expressed 
as a sum $V^{(\ell)}(\theta, \phi)=\sum_{j} V^{(\ell)}_j(\theta, \phi)$
of different $2^j$-pole potentials $V^{(\ell)}_j(\theta, \phi)$.
Although a large number of multipole potentials  might be required in general,
only a moderate number of components with different rank $j$ appear 
for up to three spins in the \drops representation of Sec.~\ref{drops_sum}~\cite{Garon15}. 
Fictitious multipole potentials sufficient to completely describe
the potential $V^{(\ell)}(\theta, \phi)$
are detailed in Table~\ref{tab:dropletlabels}: 
one has monopoles ($2^0=1$), dipoles ($2^1=2$), quadrupoles ($2^2=4$), and octupoles ($2^3=8$).
For a two-spin droplet with label ``$\{kl\}$'', only ranks $j$ of zero, one, and two occur
which correspond to fictitious monopole, dipole, and quadrupole potentials,
whereas the fully symmetric three-spin 
droplet with label ``$\{ 123\},\tau_1$'' has only rank-1 and rank-3 components
associated with dipole and octupole terms.

\subsection{Axial multipole sensors}
\label{subsec_interpret_sampl}
Based on the provided interpretation of droplet functions as fictitious multipole potentials,
the results of Sec.~\ref{drops_tomo} on how to 
experimentally measure spherical functions of spin operators
can be mapped to the analogous problem of measuring an unknown electrostatic potential.
This analogy is complete for Hermitian spin operators with
real-valued spherical functions \cite{Garon15}.
Suppose we would like to determine an unknown (real-valued) electrostatic 
potential $V(\theta, \phi)$ at a radius $\abs{\vec{r}}=1$
that is created by an object located in the interior of a unit sphere.
An electric point charge $q$ at position $\vec r$ in an electric potential $V(\vec r)$
has a  potential energy $U_{pot}(\vec r)=q V(\vec r)$.
Given an electrostatic potential $V(\theta, \phi)$,
the electrostatic potential energy of a (nonconducting) unit sphere
with the surface charge distribution $\sigma(\theta,\phi)$ is given by
\begin{equation}
\label{potA}
U_{pot}(\beta, \alpha) 
= \int_{\theta=0}^\pi \int_{\phi=0}^{2 \pi} \,  \sigma(\theta,\phi)\, V(\theta, \phi)
\,  \sin \theta \,  \mathrm{d} \theta \, \mathrm{d} \phi,
\end{equation}
which is equivalent to
$U_{pot}(\beta, \alpha) =  \langle  \sigma(\theta,\phi) \vert V(\theta, \phi)\rangle_{L^2}$
for real valued $\sigma(\theta,\phi)$
as in Eq.~\eqref{tomoprimeaa} of  Sec.~\ref{drops_tomo}.

\begin{table}[tb]
\caption{\label{tab:dropletlabels} 
Number $n$ of involved spins, components
$A^{(\ell)}$ and $f^{(\ell)}$, possible ranks $j$,
rank-$j$ contributions $A_j^{(\ell)}$ and
$f_j^{(\ell)}$, as well as $2^j$-pole potentials $V_j^{(\ell)} $
are listed for up to three spins $1/2$ and all
labels $(\ell)$; $k$ and $l$ indicate the involved spins \cite{Garon15}.}
\begin{tabular}{@{\hspace{1mm}} c @{\hspace{3mm}} l @{\hspace{2mm}} l @{\hspace{3mm}} c 
@{\hspace{3mm}}l @{\hspace{2mm}}l @{\hspace{3mm}}c @{\hspace{2mm}}c 
@{\hspace{3mm}}l @{\hspace{1mm}}} 
\\[-2mm]
\hline\hline
\\[-3mm]
$n$   & $A^{(\ell)}$ & $f^{(\ell)}$ & $j$ & $A_j^{(\ell)}$& $f_j^{(\ell)}$&  $2^j$& $2^j$-pole  & $V_j^{(\ell)} $
\\[1mm]  \hline \\[-2mm]
%-------------------------------------------------------------------------------
0&$A^{\emptyset}$ & $f^{\emptyset}$ &   0 & $A_0^{\emptyset}$ & 
$f_0^{\emptyset}$&   1& monopole &$V_0^{\emptyset}$ \\ 
\noalign{\vskip 0.5em}
1&$A^{\{ k \}}$ & $f^{\{ k \}}$ &   1 & $A_1^{\{ k \}}$&$f_1^{\{ k \}}$&   2& 
{dipole} &$V_1^{\{ k \}}$ \\    
\noalign{\vskip 0.5em}
2&$A^{\{ kl   \}}$ & $f^{\{ kl   \}}$ &   0& $A_0^{\{ kl   \}}$& $f_0^{\{ kl   \}}$  &   1&monopole &$V_0^{\{ kl   \}}$ \\   
& &  &    1 & $A_1^{\{ kl   \}}$& $f_1^{\{ kl   \}}$&   2&dipole &$V_1^{\{ kl   \}}$ \\   
& &   &   2 & $A_2^{\{ kl   \}}$& $f_2^{\{ kl   \}}$&  4& quadrupole  &$V_2^{\{ kl   \}}$\\ 
\noalign{\vskip 0.5em}
3&$A^{\tau_1}$ &  $f^{\tau_1}$ &  1 &  $A_1^{\tau_1}$&  $f_1^{\tau_1}$&  2& dipole  &$V_1^{\tau_1}$ \\   
& &  &    3&  $A_3^{\tau_1}$&  $f_3^{\tau_1}$ & 8& octupole &$V_3^{\tau_1}$ \\   
 \noalign{\vskip 0.5em}
&$A^{\tau_2}$ &  $f^{\tau_2}$ &   1 &  $A_1^{\tau_2}$&  $f_1^{\tau_2}$&   2&dipole  &$V_1^{\tau_2}$ \\   
&&   &   2 &  $A_2^{\tau_2}$ &  $f_2^{\tau_2}$ & 4& quadrupole &$V_2^{\tau_2}$  \\ 
\noalign{\vskip 0.5em}  
&$A^{\tau_3}$ & $f^{\tau_3}$ &   1&  $A_1^{\tau_3}$&  $f_1^{\tau_3}$&   2&dipole & $V_1^{\tau_3}$ \\   
&&   &   2&  $A_2^{\tau_3}$&  $f_2^{\tau_3}$&  4&quadrupole  &$V_2^{\tau_3}$ \\   
\noalign{\vskip 0.5em}
&$A^{\tau_4}$ &$f^{\tau_4}$ &    0&  $A_0^{\tau_4}$&  $f_0^{\tau_4}$&   1&monopole &$V_0^{\tau_4}$
%-------------------------------------------------------------------------------
\\[1mm]  \hline \hline  \\[-2mm]
\end{tabular}
\end{table}

The $2^j$-pole components $V_j(\theta,\phi)$ of an unknown 
multipole potential $V(\theta, \phi)$ can be sampled by a set of axial $2^j$-pole sensors, 
each consisting of a charge distribution 
$\sigma_{j0}(\theta,\phi)$ proportional to the axial spherical harmonics $Y_{j0}(\theta, \phi)$.
Each individual sample $V_j(\beta,\alpha)$ can be determined using
Eq.~\eqref{potA}  by measuring the potential energy 
\begin{equation*}
V_j(\beta, \alpha)\propto  U_{pot}(\beta,\alpha) =  \langle {R}_{\alpha \beta} 
\, \sigma_{j0}(\theta, \phi) \vert   V(\theta, \phi)\rangle
\end{equation*}
of the axial $2^j$-pole sensor rotated by the polar angle $\beta$ around 
the $y$ axis followed by a rotation by the azimuthal angle $\alpha$ around the $z$ axis. 
The full electrostatic $2^j$-pole potential $V_j(\theta, \phi)$ can be recovered
by systematically incrementing $\beta$ and $\alpha$.
In summary, the analogy between real-valued spherical functions
and multipole potentials helps to better understand our results of Sec.~\ref{drops_tomo}
on the measurement of spherical functions. It can also
be extended in a straight-forward manner to non-Hermitian spin operators 
by considering complex, fictitious multipole potentials.

%-----------------------------------------------------------------------------------------------------
%---- Results
%-----------------------------------------------------------------------------------------------------

\section{Summary of NMR experiments \label{exp_res}}

\begin{figure}[tb]
\centering
\includegraphics{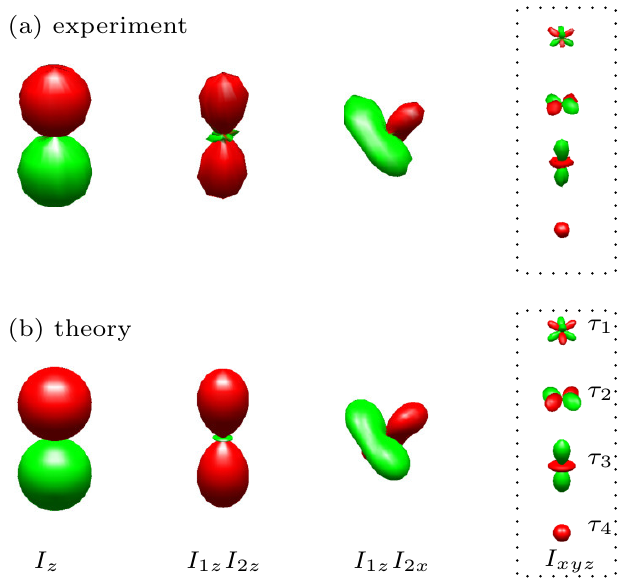}
\caption{(Color online) Spherical functions from (a) experiment and (b) theory; 
$I_{xyz}:=I_{1x}I_{2y}I_{3z}$ splits into 
$f^{\tau_1}$, $f^{\tau_2}$, $f^{\tau_3}$, and $f^{\tau_4}$ (see\
Fig.~\ref{fig:mapping_new}) \cite{Note\thecolors}. Further examples are shown
in Figs.~\ref{over_b} and \ref{over_b2}.
\label{over_a}}
\end{figure}

Building on the previous sections, we demonstrate the Wigner
tomography of various prepared density-matrix components in spin-$1/2$ 
systems using nuclear magnetic resonance. Experimental details
are deferred to Secs.~\ref{theory} and \ref{exp_impl} where
the precise experimental scheme and its implementation on a spectrometer
are discussed. The experiments were performed on
one-, two-, and three-spin systems. 
The shapes of the spherical functions are recovered
for the  prepared Cartesian products operators listed in Table~\ref{tab:errors},
where also their respective experimental reconstruction errors are given.
Experimental and theoretical results for the reconstruction 
are visually compared for four examples in Fig.~\ref{over_a}.
For the rightmost example of $4I_{1x}I_{2y}I_{3z}$ in Fig.~\ref{over_a}, multiple droplets 
corresponding to different permutation symmetries are necessary to completely describe 
the quantum operator,  as outlined in Sec.~\ref{drops_sum} (see Fig.~\ref{fig:mapping_new}
and \cite{Garon15}).

\begin{table}[tb]
\caption{Root-mean-square error of the experimental 
reconstruction for the prepared Cartesian product operators
on one-, two-, and three-spin systems;
$I_{abc}:=I_{1a}I_{2b}I_{3c}$.
\label{tab:errors}}
\begin{tabular}{@{\hspace{1mm}} c @{\hspace{3mm}} l @{\hspace{3mm}} c 
@{\hspace{8mm}} c @{\hspace{3mm}} l @{\hspace{3mm}} c @{\hspace{1mm}}} 
\\[-2mm]
\hline\hline
\\[-3mm]
spins & prod. op. & error & spins & prod. op. & error
\\[1mm]  \hline \\[-2mm]
%-------------------------------------------------------------------------------
1 & $I_x$ & 0.0519 & 3  & $4I_{xxx}$ & 0.0797 \\
& $I_y$ & 0.1187 & & $4I_{yyy}$ & 0.0509 \\
& $I_z$ & 0.0356 & & $4I_{xyz}$ & 0.0617 \\
\noalign{\vskip 0.5em}
2 & $2I_{1x}I_{2x}$ & 0.0173 & & $4I_{xyy}$ & 0.0395 \\
&  $2I_{1y}I_{2y}$ & 0.0154 & & $4I_{yxy}$ & 0.0476 \\
&  $2I_{1z}I_{2z}$ & 0.0850 & & $4I_{yyx}$ & 0.0954 \\
&  $2I_{1x}I_{2y}$ & 0.0487 & & $4I_{xxy}$ & 0.0587 \\
&  $2I_{1y}I_{2x}$ & 0.0152 & & $4I_{xyx}$ & 0.0638 \\
&  $2I_{1z}I_{2x}$ & 0.0319 & & $4I_{yxx}$ & 0.0692 \\
%-------------------------------------------------------------------------------
\hline \hline 
\end{tabular}
\end{table}

\section{Temporal Averaging\label{temporal}}

\begin{figure}[tb]
\centering
\includegraphics{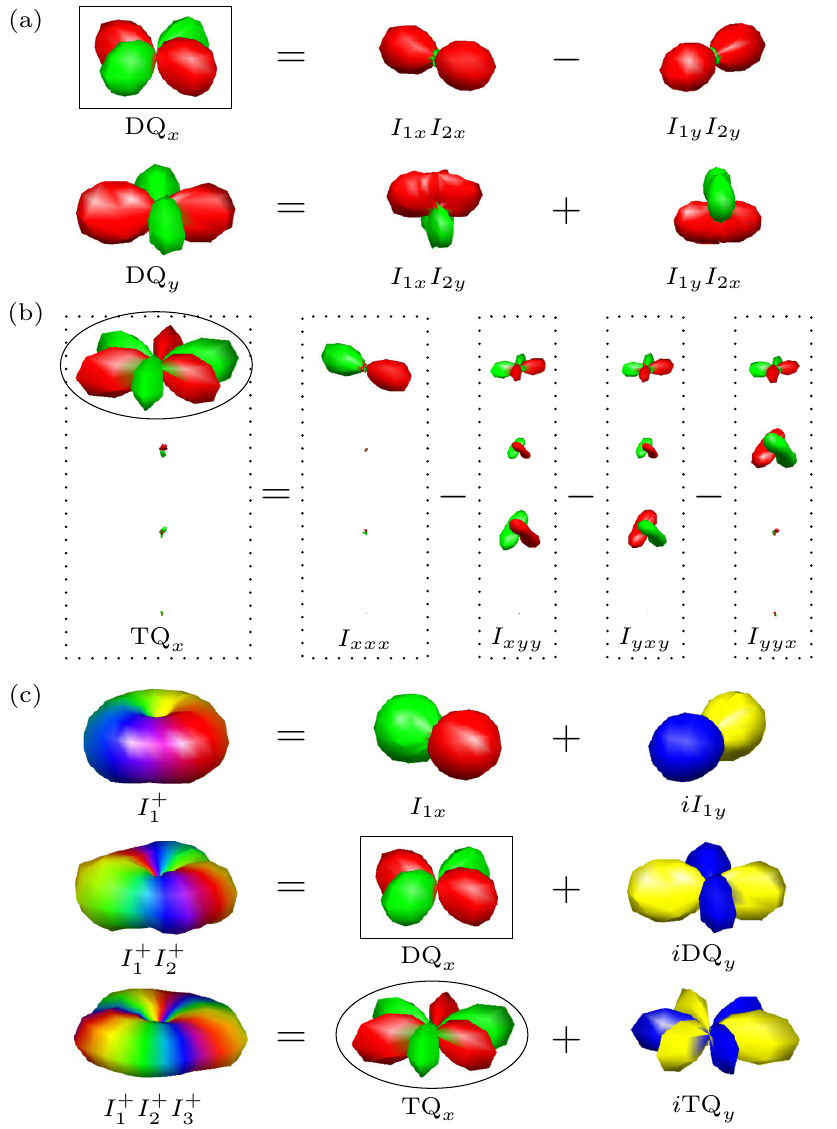}
\caption{(Color online) Temporal averaging:
decomposition of 
(a) $\text{DQ}_x$ [box, see also panel (c)], $\text{DQ}_y$, and (b) 
$\text{TQ}_x$ [$f^{\tau_1}$ ellipse, see also panel (c)],
$I_{abc}:=I_{1a}I_{2b}I_{3c}$.
(c) non-Hermitian operators $I_1^+$, $I_1^+ I_2^+$, and $I_1^+ I_2^+ I_3^+$ 
as complex linear combination of Hermitian ones (see Fig.~\ref{over_a}).
\label{over_b}}
\end{figure}

\begin{figure}[tb]
\centering
\includegraphics{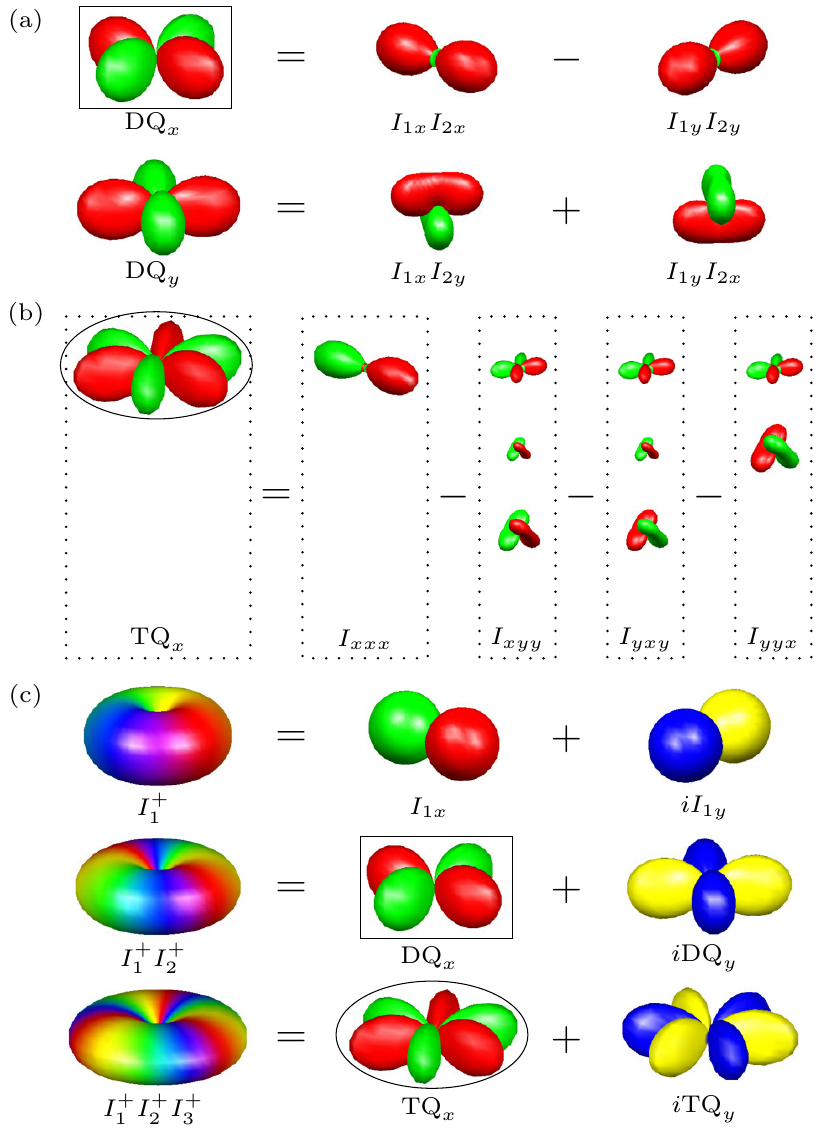}
\caption{(Color online) Theoretical predictions corresponding to 
Fig.~\ref{over_b}; $I_{abc}:=I_{1a}I_{2b}I_{3c}$.
\label{over_b2}}
\end{figure}

Re-using our experimental data as summarized 
in Sec.~\ref{exp_res}, we can also highlight how temporal averaging \cite{PhysRevLett.80.3408}
is used to emulate the preparation of quantum operators.
The direct experimental preparation of Hermitian operators would be also
possible, but we have chosen temporal averaging for its
simplicity and convenience. The experimental values are shown in Fig.~\ref{over_b},
while the corresponding theoretical predictions are given in
Fig.~\ref{over_b2}.
The Cartesian operators
$2I_{1x}I_{2x}$, $2I_{1y}I_{2y}$, $2I_{1x}I_{2y}$, and 
$2I_{1y}I_{2x}$ had been sequentially measured
and are now combined in 
Fig.~\ref{over_b}(a) to form the 
double quantum operators 
\begin{equation*}
\text{DQ}_x := I_{1x}I_{2x} - I_{1y}I_{2y} \, \text{ and } \,
\text{DQ}_y := I_{1x}I_{2y} + I_{1y}I_{2x}.
\end{equation*}
Their characteristic shapes reflect the fact that they have 
coherence order $\abs{p}=2$ and are invariant under
non-selective rotations around the $z$ axis by an integer multiple of $180$ deg
\cite{Garon15}.
This is in contrast to
single-quantum operators such as the linear operators $I_x$
or $I_y$ [see\ \ref{over_b}(c)] 
or the bilinear operator  $2 I_{1z} I_{2x}$ 
[see\ Fig.~\ref{over_a}]  which 
are only invariant under non-selective rotations around the $z$ 
axis by an integer multiple of $360$ deg \cite{EBW87}.

In general, $\abs{p}$-quantum operators 
are invariant under non-selective rotations around the $z$ axis by an integer multiple of 
$360/\abs{p}$ deg, and their spherical functions illustrate this symmetry.
Figure~\ref{over_b}(b) exemplifies the invariance under $120$ deg rotations around 
the $z$ axis in the case of $\abs{p}=3$ for the triple-quantum
operator 
\begin{align*}
\text{TQ}_x &:= I_{1x}I_{2x}I_{3x}{-}I_{1x}I_{2y}I_{3y}{-}I_{1y}I_{2x}I_{3y}
{-}I_{1y}I_{2y}I_{3x},\\
\intertext{similar to the case of}
\text{TQ}_y &:=I_{1y}I_{2x}I_{3x}{+}I_{1x}I_{2y}I_{3x} 
{+}I_{1x}I_{2x}I_{3y}{-}I_{1y}I_{2y}I_{3y}.
\end{align*}
Up to experimental imperfections, 
only the spherical function $f^{\tau_1}$ 
contributes to the operator $\text{TQ}_x$.

Finally, we also consider temporal averaging 
for non-Hermitian operators which obviously cannot 
be directly prepared in experiments (see also Sec.~\ref{theory}).
Figure~\ref{over_b}(c) presents the non-Hermitian operators
$I_1^{+} = I_{1x}+iI_{2y}$, 
$I_1^{+}I_2^{+} = \text{DQ}_x + i\, \text{DQ}_y$, and
$I_1^{+}I_2^{+}I_3^{+}  = \text{TQ}_x + i\, \text{TQ}_y$.
These operators have the respective
coherence orders $p$ of $1$, $2$, and $3$ which results
in donut-shaped spherical functions the colors of which
cycle through one, two, or three rainbows \cite{Garon15}.

\section{Wigner tomography using NMR \label{theory}}

\begin{figure}[tb]
\centering
\includegraphics{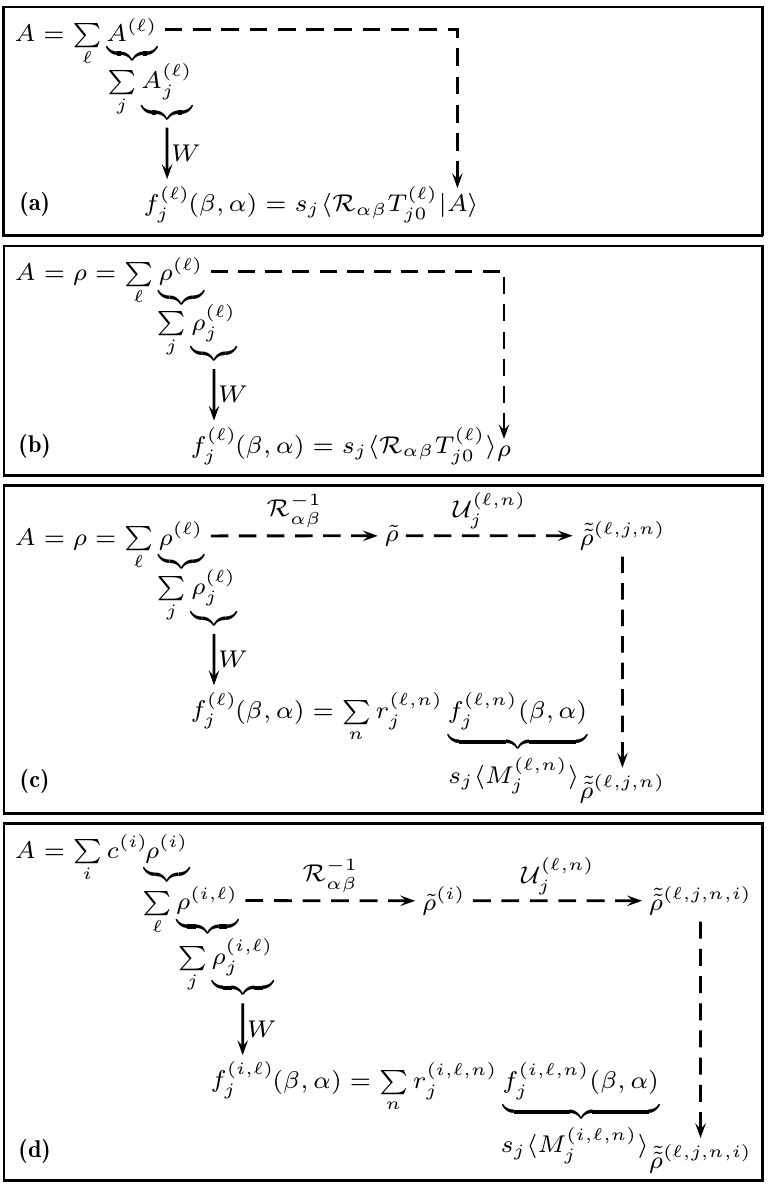}
\caption{Sampling schemes for  
spherical functions using (a) scalar products of 
rotated axial tensor operators and $A$ 
[Result~\ref{result2}, Eq.~\eqref{tomoprime}].
(b) Expectation values of rotated axial tensor operators
w.r.t.\ $\rho$ [Result~\ref{result2}, Eq.~\eqref{exptomoprime}].
 (c) Experimentally measurable expectation values as in panel (b) 
[Result~\ref{result3}, Eqs.~\eqref{tomozzz}-\eqref{tomozzzc}].
(d) non-Hermitian operators
averaged as complex combinations of Hermitian 
terms  $\rho^{(i)}$ [see \ Eq.~\eqref{exptomoprimexxx}]. 
The Wigner transformation $W$ maps an operator to its spherical function.
\label{Tomo_Fig}}
\end{figure}

We complement our results in Sec.~\ref{drops_tomo} and describe 
the experimental scheme for an NMR-based implementation of our
Wigner tomography. Recall that Eq.~\eqref{tomoprime} of Result~\ref{result2} 
provides an approach for measuring an arbitrary operator. This can be translated 
into the diagram of Fig.~\ref{Tomo_Fig}(a): The operator $A$ is decomposed into
its components $A_j^{(\ell)}$ which are mapped by the Wigner transformation $W$
to spherical samples $f_j^{(\ell)}(\beta,\alpha)$. The spherical samples
can be recovered using Eq.~\eqref{tomoprime}. Very similarly, Fig.~\ref{Tomo_Fig}(b)
depicts the equivalent measurement procedure for density matrices which
relies on Eq.~\eqref{exptomoprime} of Result~\ref{result2}.

Before proceeding to the NMR-based scheme, it is important 
to emphasize that only traceless operators can be measured 
in NMR experiments \cite{EBW87}, which rules out the 
identity component of a density matrix. However, the traceless part of 
a density matrix covers most important features, and it is
sufficient  to calculate the time evolution and 
all relevant expectation values. Hence, we will 
consider in the following only the  traceless part which is 
for simplicity also denoted by $\rho$.

Further complications arise from the fact that signatures of 
Cartesian product operators \cite{EBW87}
that contain only a single transverse Cartesian operator $I_{ka}$ with $a\in \{ x, \ y\}$ can be measured 
\emph{directly}; examples are $I_{ka}$, $2I_{ka}I_{lz}$, and $4I_{ka}I_{lz} I_{mz}$.
This complication can be resolved in two steps:  
Firstly, any traceless operator can be decomposed into (Hermitian) Cartesian product operators
$C^{(\ell,n)}_{j}$.
This decomposition of relevant axial tensors 
\begin{equation}
\label{measureops}
T^{(\ell)}_{j0}=  
 \sum_n r^{(\ell,n)}_{j} C^{(\ell,n)}_{j}
\end{equation}
with respect to real coefficients $r^{(\ell,n)}_{j}$
is given in Table~\ref{tab:axTens}.
For example,
the axial tensor operator $T_{10}^{\{ k\}}$ acting on the $k{\rm th}$ spin 
decomposes directly into
the Cartesian product operator 
$C_1^{(\{k\},1)}=I_{kz}$
with  the coefficient $r_1^{(\{k\},1)}=\sqrt{2}$.

Secondly, the Cartesian product operators have to be transformed
into NMR-measurable ones
\begin{equation}
\label{detopa}
M^{(\ell,n)}_{j}= \mathcal{U}^{(\ell,n)}_{j} C^{(\ell,n)}_{j} := {U}^{(\ell,n)}_{j} C^{(\ell,n)}_{j} {U}^{(\ell,n)\dagger}_{j}.
\end{equation}
The unitary operators ${U}^{(\ell,n)}_{j}$
can be experimentally realized
using radio-frequency pulses and evolution periods under couplings
as discussed in Sec.~\ref{exp_proto} and 
their explicit form is detailed in 
App.~\ref{prep_detec}. Combining both steps leads to an
\emph{indirect} approach for measuring spherical functions of density operators, 
as schematically outlined in Fig.~\ref{Tomo_Fig}(c). 
The density matrix is equivalently rotated inversely
in contrast to  Fig.~\ref{Tomo_Fig}(a)-(b) where the axial tensor operator $T_{j0}^{(\ell)}$
is rotated. The complete measurement scheme is formalized 
along the lines of Result~\ref{result2}:

\begin{result}
\label{result3}
Consider a density operator $\rho$ which is represented by a set of
spherical functions $f^{(\ell)}(\theta,\phi)=\sum_{j \in J(\ell)} f_{j}^{(\ell)}(\theta,\phi)$.
For each label $\ell$, the rank-$j$  component $f_{j}^{(\ell)}(\beta,\alpha)$
can be  measured for arbitrary angles $\beta$ and $\alpha$
by determining the expectation values 
\begin{align}
\label{tomozzz}
f^{(\ell)}_j(\beta, \alpha)
& = s_j   \sum_n \hskip 0.1em r^{(\ell,n)}_{j} \hskip 0.1em \langle M^{(\ell,n)}_{j} 
\rangle_{\tilde{\tilde \rho}^{(\ell,j,n)}}
\intertext{of suitable operators $M^{(\ell,n)}_{j}$ as in Eq.~\eqref{detopa}, where} 
\label{tomozzzb}
{\tilde{\tilde \rho}}^{(\ell,j,n)}
& = \mathcal{U}^{(\ell,n)}_{j}  \, \tilde \rho
={ U}^{(\ell,n)}_{j}  \, \tilde \rho \, { U}^{(\ell,n)\dagger}_{j}, \\
\label{tomozzzc}
{{\tilde \rho}}
& = {\cal R}^{-1}_{\alpha \beta} \, \rho
={ \mathfrak{R}}^{-1}_{\alpha \beta} \, \rho \, { \mathfrak{R}}_{\alpha \beta},
\end{align}
and 
${ \mathfrak{R}}_{\alpha \beta}=\exp(- i \alpha \sum_{k=1}^n I_{kz}) \exp(- i \beta \sum_{k=1}^n I_{ky})$.
\end{result}
A detailed derivation of Eq.~\eqref{tomozzz} is provided in Appendix \ref{sec_drops_NMR_spec}. 
In summary, the rank-$j$ components $f^{(\ell)}_j$ of spherical functions $f^{(\ell)}$ 
representing the density matrix $\rho$ can be sampled in NMR experiments by 
transforming the density operator $\rho$ to the states $\tilde {\tilde \rho}^{(n)}$ and 
then measuring a set of expectation values of suitable
operators $\langle M^{(\ell,n)}_{j}  \rangle_{\tilde{\tilde \rho}^{(\ell,j,n)}}$. The explicit form 
of the Cartesian operators $C^{(\ell,n)}_{j}$, their 
NMR-measurable forms  $ M^{(\ell,n)}_{j}$, and 
the transformations ${ U}^{(\ell,n)}_{j}$ for up to three
spins is given in App.~\ref{prep_detec}.

The approach of Result~\ref{result3} can be extended to non-Hermitian operators,
even though these cannot be prepared directly in an experiment. We apply
temporal averaging \cite{PhysRevLett.80.3408} as already discussed in Sec.~\ref{temporal}.
Any operator $A$ can be expressed as a complex linear combination
$A=  \sum_i c^{(i)}  \rho^{(i)}$ of Hermitian operators $\rho^{(i)}$.
As the \drops representation is linear, we can sample the traceless 
part of any operator $A$ using the spherical functions
\begin{align}
\label{exptomoprimexxx}
f^{(\ell)}_j(\beta, \alpha)
&=   \sum_i c^{(i)} f^{(i,\ell)}_j(\beta, \alpha) \nonumber \\
&= \sum_i c^{(i)}  \,  s_j  \, \langle  {\cal R}_{\alpha \beta} T^{(\ell)}_{j0} \rangle_{\rho^{(i)} } \\
&=\sum_i c^{(i)}  \,  s_j  \sum_n \, r^{(\ell,n)}_{j} 
\, \langle M^{(\ell,n)}_{j} \, \rangle_{\tilde{\tilde \rho}^{(\ell,j,n,i)}}. \nonumber
\end{align}
In an experiment,  temporal averaging 
of Hermitian operators $\rho^{(i)}$
can be implemented by sequentially measuring spherical functions for each operator
$\rho^{(i)} $ and linearly combining the results
$f^{(i,\ell)}_j(\beta, \alpha)=s_j  \, \langle  {\cal R}_{\alpha \beta} T^{(\ell)}_{j0} \rangle_{\rho^{(i)} } 
= s_j  \sum_n \, r^{(\ell,n)}_{j} \hskip 0.1em \langle M^{(\ell,n)}_{j} 
\, \rangle_{\tilde{\tilde \rho}^{(\ell,j,n,i)}}$
as illustrated in Fig.~\ref{Tomo_Fig}(d).

\begin{table}[tb]
\caption{Axial tensor operators $T^{(\ell)}_{j0}$ 
and their decomposition into Cartesian product operators $C^{(\ell,n)}_{j}$  for three spins~\cite{Garon15}.
\label{tab:axTens} }
\begin{tabular}{@{\hspace{1mm}} l @{\hspace{3mm}} l @{\hspace{1mm}}} 
\\[-2mm]
\hline\hline
\\[-3mm]
$T_{j0}^{(\ell)}$ & $\sum_n r_j^{(\ell,n)} C_j^{(\ell,n)}$ 
\\[1mm]  \hline \\[-2mm]
%-------------------------------------------------------------------------------
$T_{10}^{\{k\}}$ & $\sqrt{2}I_{kz}$\\
\noalign{\vskip 0.5em}
$T_{00}^{\{kl\}}$ & $(2I_{kx}I_{lx}+2I_{ky}I_{ly}+2I_{kz}I_{lz})/\sqrt{3}$\\
$T_{10}^{\{kl\}}$ & $(2I_{kx}I_{ly}-2I_{ky}I_{lx})/\sqrt{2}$\\
$T_{20}^{\{kl\}}$ & $[-2I_{kx}I_{lx}-2I_{ky}I_{ly}+2(2I_{kz}I_{lz})]/\sqrt{6}$ \\
\noalign{\vskip 0.5em}
$T_{10}^{\tau_1}$ & $\sqrt{8}(I_{xxz}{+}I_{xzx}{+}I_{zxx} {+}I_{yyz}{+}I_{yzy}{+} I_{zyy}{+}3 I_{zzz})/\sqrt{15}$\\
$T_{30}^{\tau_1}$ & ${-}2(I_{xxz}{+}I_{xzx}{+}I_{zxx}{+}I_{yyz}{+}I_{yzy}{+}I_{zyy}{-}2I_{zzz})/\sqrt{5}$\\
$T_{10}^{\tau_2}$ & $\sqrt{2}[ -2(I_{xxz}{+}I_{yyz}){+}I_{zxx}{+}I_{xzx}{+}I_{zyy}{+}I_{yzy}]/\sqrt{3}$\\
$T_{20}^{\tau_2}$ & $\sqrt{2} ( I_{yzx}{+}I_{zyx}{-}I_{xzy}{-}I_{zxy})$\\
$T_{10}^{\tau_3}$ & $\sqrt{2} ( I_{zxx}{-}I_{xzx}{+}I_{zyy}{-}I_{yzy})$\\
$T_{20}^{\tau_3}$ & $\sqrt{2}[ -2(I_{xyz}{-}I_{yxz}){+}I_{zxy}{-}I_{xzy}{+}I_{yzx}{-}I_{zyx}]/\sqrt{3}$\\
$T_{00}^{\tau_4}$ & $2(I_{xyz}{-}I_{xzy}{-}I_{yxz}{+}I_{yzx}{+}I_{zxy}{-}I_{zyx})/{\sqrt{3}}$
%-------------------------------------------------------------------------------
\\[1mm]  \hline \hline  \\[-2mm]
\end{tabular}
\end{table}

%-----------------------------------------------------------------------------------------------------%
%---- Experiment
%-----------------------------------------------------------------------------------------------------

\section{Experimental Implementation\label{exp_impl}}

After outlining the experimental scheme for Wigner tomography
in Sec.~\ref{theory}, we present now the details for the experimental implementation
which results in the spherical functions in Figs.~\ref{over_a}-\ref{over_b}.
We start by describing the molecules and experimental setting 
in Sec.~\ref{sec_mol}. We continue in Sec.~\ref{exp_proto} with the experimental protocol,
and we finally discuss experimental errors in Sec.~\ref{sec_exp_error}.

%-----------------------------------------------------------------------------------------------------
%---- molecules
\subsection{Molecules and experimental setting\label{sec_mol}}

\begin{figure}[tb]
\includegraphics{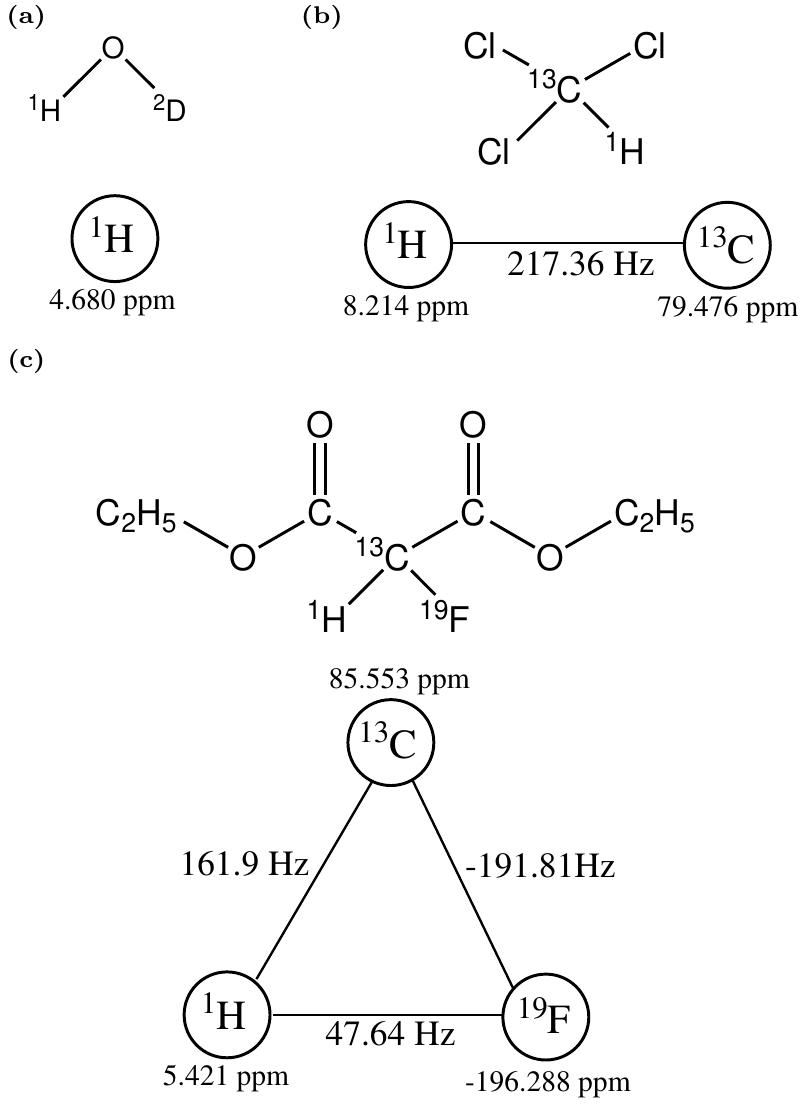}
\caption{Molecules 
(a) HDO, (b) chloroform, and (c) 2-$^{13}$C-2-fluoromalonic-acid-diethyl-ester
used in experiments with their schematic spin systems and coupling topologies;
individual spins are labeled by chemical shifts (in parts per million);
heteronuclear $J$ couplings (lines) are
labeled by coupling constants $J_{kl}$ (in hertz).
\label{mol}}
\end{figure}

In order to simplify the experiments, the linear and bilinear Cartesian product 
operators have been prepared and measured using 
respective single-spin and two-spin samples: The single-spin sample
was prepared by dissolving
5\% H$_2$O in D$_2$O, which resulted in a sample containing about
10\% HDO, i.e.\  water molecules in which one of the $^1$H spins is replaced by 
deuterium ($^2$D) [see\ Fig.~\ref{mol}(a)]. In case of two spins,
we have used a 10\% sample of chloroform dissolved in fully deuterated DMSO-d6, 
where the $^1$H spin  and the $^{13}$C spin of each chloroform molecule 
form a system consisting of two coupled heteronuclear spins $1/2$ [see\ Fig.~\ref{mol}(b)].
A three-spin sample consisting 
of 2-$^{13}$C-2-fluoromalonic-acid-diethyl-ester dissolved in CD$_3$CN
[see Fig.~\ref{mol}(c)]
was utilized  for the preparation and  reconstruction of trilinear operators. 
All liquid samples were measured in 5 mm Shigemi NMR tubes at room temperature (298 K)
in  a 14.1 T magnet
using a Bruker Avance III 600 spectrometer.

%-----------------------------------------------------------------------------------------------------
%---- preparations
\subsection{Experimental protocol\label{exp_proto}}

\begin{figure}[tb]
\includegraphics{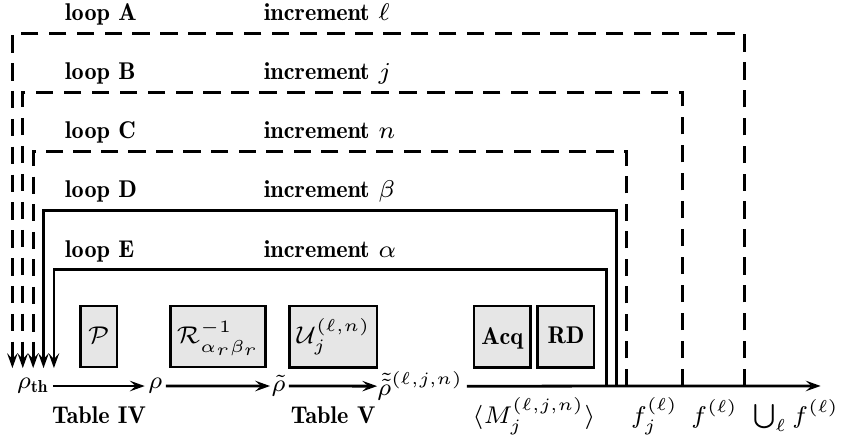}
\caption{Tomography scheme proposed by Result~\ref{result3}; 
note that $\langle M_j^{(\ell,j,n)}\rangle := \langle M_j^{(\ell,j,n)}\rangle_{\tilde{\tilde{\rho}}^{(\ell,j,n)}}$.
\label{tomo_exp}}
\end{figure}

Our experimental protocol is composed of 
five main building blocks (see Fig.~\ref{tomo_exp}).
In the first block $\mathcal{P}$, the desired density operator $\rho$ is
prepared starting from the initial thermal equilibrium density operator
which in the high-temperature limit is 
proportional to \cite{EBW87}
${\rho_{th}} = 
\sum_{k=1}^{N}\gamma_k{I_{kz}}$,
where $\gamma_k$ denotes the gyromagnetic ratio of the $k$th nuclear spin. 
This requires
unitary transformations which are created by pulses and evolution periods under the 
effect of couplings and frequency offsets as well as non-unitary transformations which are
implemented by pulsed magnetic-field gradients. The explicit pulse sequences 
are discussed in App.~\ref{prep_detec}.
Table~\ref{tab:errors} in Sec.~\ref{exp_res} summarizes
all the different Cartesian product operators which have been experimentally prepared.

The second block consists of the rotation $\mathcal{R}^{-1}_{\alpha_r\beta_r}$
which rotates the prepared  density operator $\rho$ 
into $\tilde{\rho}$
in order to probe the corresponding spherical functions 
$f_j^{(\ell)}(\beta_r, \alpha_r)$ for different  polar angles $\beta_r$ and azimuthal angles $\alpha_r$
using axial tensor operators, i.e., \ axial multipole sensors)
(see\ Result~\ref{result3}). The rotation $\mathcal{R}^{-1}_{\alpha_r \beta_r}$ 
is implemented by rf pulses $[\beta_r]_{\alpha_r-\pi/2}$
with flip angle $\beta_r$ and phase $(\alpha_r-\pi/2)$
which are simultaneously applied to all spins.

The unitary transformations $\mathcal{U}_j^{(\ell,n)}$
[see\ Eq.~\eqref{tomozzzb} of Result~\ref{result3}]
are applied in the third block in order to transform the density matrix $\tilde \rho$ into
directly detectable Cartesian product operators
for the various linear, bilinear, 
and trilinear operators (see\ Table~\ref{tab:axTens} in Sec.~\ref{theory}).
The specific experimental implementation of the unitary transformations $\mathcal{U}^{(\ell,n)}_{j}$ 
consists of rf pulses with flip angle $\pi/2$
as detailed in App.~\ref{prep_detec}.

In the fourth block, the NMR signal is measured in an acquisition period Acq
which has a duration of 5.7 ms (one spin), 
11.4 ms (two spins), and 14.9 ms (three spins). In the last block,
a relaxation delay RD with a duration of 
7 s (one spin), 10 s (two spins), and 15 s (three spins)
recovers the initial equilibrium state $\rho_{th}$,

In the tomography experiment, all blocks are repeated multiple times 
(see Fig.~\ref{tomo_exp}).
The outer loop A runs over all possible droplets $\ell \in L$. Loop B  runs over all 
ranks $j$ contributing to the droplet $\ell$. Loop C 
cycles through
all Cartesian product operators  $C^{(\ell,n)}_j$ [see\ Eq.~\eqref{measureops}] appearing
in the decomposition of the axial tensor operator $T^{(\ell)}_{j0}$ (see\ Table~\ref{tab:axTens} in 
Sec.~\ref{theory}). Finally, the discretized angles 
$\beta_r\in \{0,  15, 30, \ldots , 180\}$ and 
$\alpha_r\in \{0,  15, 30, \ldots , 360\}$ (both in degrees)
are incremented in the innermost loops D and E.
Although not explicitly indicated in Fig.~\ref{tomo_exp}, one further loop is necessary for
a temporal averaging scheme [see\  Eq.~\eqref{exptomoprimexxx}].

\begin{figure}[tb]
\includegraphics{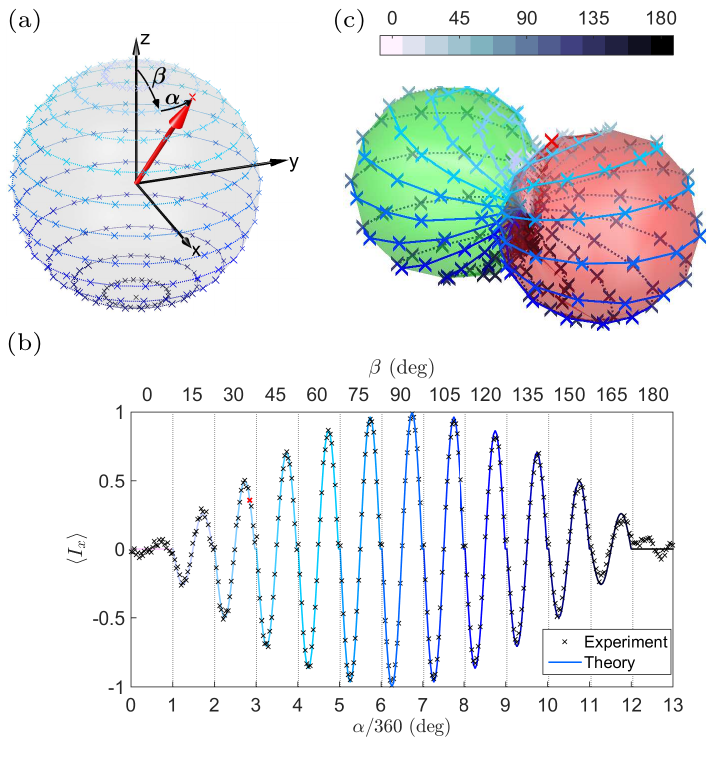} 
\caption{(Color online) Reconstruction of a spherical function from experimental samples
$f^{(\ell)}(\beta_r,\alpha_r)$.
(a) Samples (crosses)
with different polar angles $\beta_r \in \{0,180\}$ in degrees
(circles colored
by latitude)
and phases $\alpha_r \in \{0,360\}$ in degrees
acquired using rf-pulses $[\beta_r]_{\alpha_r - \pi/2}$.
(b) Predicted expectation values 
$\langle M_1^{(1,1)} \rangle_{\tilde{\tilde{\rho}}^{(\ell,j,n)}} = \langle I_{1x} \rangle$ 
depending on
a discrete set of polar angles
$\beta \in\{0, 15, \ldots, 180\}$ 
in degrees
and 
a continuous set of azimuthal angles
$\alpha$.
(c) Smooth surface interpolated
from indiviual
samples, 
distance from the origin given by $f^{(\ell)}(\beta_r,\alpha_r)$
whose phase determines the color of the surface (see\ Fig.~\ref{fig:mapping_new}).
\label{fig_reconstr}}
\end{figure}

The whole protocol allows us to 
determine expectation 
values $\langle M_j^{(\ell,n)} \rangle_{\tilde{\tilde{\rho}}^{(\ell,j,n)}}$,
which are normalized and range between 1 and $-1$.
As illustrated in Fig.~\ref{fig_reconstr}, the spherical functions
$f^{(\ell)}(\theta,\phi)$
can be reconstructed
by plotting the spherical samples 
$f^{(\ell)}(\beta_r,\alpha_r)$ for all angles
 $\beta_r \in \{0,\dots,\pi\}$ 
and $\alpha_r \in \{0,\dots,2\pi\}$ at a distance $\vert f^{(\ell)}(\beta_r,\alpha_r) \vert$ from the origin. 
The phase 
$\varphi^{(\ell)}   (\beta_r, \alpha_r)= {\rm atan} ({\rm Im} 
\{   f^{(\ell)}(\beta_r, \alpha_r)   \}   / {\rm Re} \{   f^{(\ell)}(\beta_r, \alpha_r)   \}     )$ 
is color coded. For example, the spherical function of the Hermitian operator $I_{1x}$ is given by a
real function, and the positive and 
negative values of $f^{(\ell)}(\theta,\phi) $ are indicated by the colors red  (dark gray) and green (light gray), 
respectively.

\subsection{Experimental errors\label{sec_exp_error}}
A reasonable match between the experimentally reconstructed and 
theoretical predicted spherical functions is found in Figs.~\ref{over_a}-\ref{over_b2}. 
Deviations are attributed to experimental imperfections, such as the 
finite experimental signal-to-noise ratio, finite accuracy of pulse calibration, 
$B_0$ and $B_1$ inhomogeneity~\cite{levitt_book, EBW87}, 
pulse shape distortions due to the amplifiers and the finite bandwidth of the resonator~\cite{transient},  
relaxation losses during the preparation and detection blocks,
partial saturation of the signal due to a finite relaxation period between scans,
radiation damping effects~\cite{CMR:CMR1}, and
truncation effects in the automated integration and comparison of the spectra.
We quantify these deviations by 
the root-mean-square difference between experiment and theory
averaged over all measured angles $\beta_r$ and $\alpha_r$.
The resulting errors for the prepared and measured Cartesian product operators 
are summarized in Table~\ref{tab:errors} of Sec.~\ref{theory}.
In order to minimize the effects of field 
inhomogeneities, Shigemi tubes were used in the experiments in order to reduce the sample volume. 
We have tested replacing 
the simple rectangular pulses $[\beta_r]_{\alpha_r-\pi/2}$ 
in the implementation of  the rotations
$\mathcal{R}^{-1}_{\alpha_r \beta_r}$
with  composite pulses \cite{Levitt2005}, which,
however, did not result in an improved performance.

\section{Conclusion\label{conclusion}}

We have theoretically developed and experimentally demonstrated a Wigner tomography scheme the mapping of which 
multi-spin operators and spherical functions is based on~\cite{Garon15}.
Our approach reconstructs the relevant spherical functions
by measuring expectation values of rotated axial tensor operators, i.e.\ 
axial multipole sensors. It is universally applicable and 
not restricted to NMR methodologies or particles with spin $1/2$.
A reasonable match between theoretical predictions 
and NMR experiments was found. 

Our theoretical analysis provides a simple physical interpretation of the individual 
spherical functions in terms of fictitious multipole potentials
which can be sampled locally. The objective was to 
experimentally recover the three-dimensional shapes of the spherical
functions for each of the prepared operators. In particular, 
we have not used any \textit{a priori} information on what shapes
to expect. A large number of 
sampling points is necessary to recover the shapes in sufficient detail.
As the simple rectangular grid of sampling points
in the space of polar 
and azimuthal angles used here is highly anisotropic
(i.e.\ more densely concentrated at the poles), 
a straight-forward improvement is to choose 
more isotropic sampling strategies such as Lebedev grids~\cite{leb75, leb76, leb99}.
Moreover, one could use interpolation methods or even adaptive sampling schemes
which increase the sampling density in areas where the spherical function varies 
more strongly. A more quantitative analysis of the complexity of
our proposed tomography approach in terms
of the number of individual measurements could rely first on a detailed 
empirical account on how the precision measured by the fidelity depends on 
the sampling strategy and its density. From the theoretical side, it is clear that
the number of measurements for any complete tomography of a quantum state
will scale exponentially in the number of qubits (or spins). However, relevant
information can be reconstructed in our approach even from a subset 
of the droplets as discussed under (c) in Sec.~\ref{drops_sum}.
A  more accurate understanding of concrete sampling schemes 
and their optimization building on 
our reconstruction method will have to be addressed in future work.

Beyond the optical tomographic methods mentioned at the start of the 
introduction, tomographic approaches play obviously an important part in 
almost any experiment  in quantum information or quantum physics 
in general. We will now shortly discuss some selected results from the literature.
The proposed reconstruction procedure for spherical functions
can be directly extended to other types of spherical representations
such as multipole operators \cite{Garon15} or the 
so-called PROPS representation which 
is based on products of single-spin representations \cite{koczor2016}.
Our scheme can also be compared 
to
\cite{tilma2016} which introduces Wigner functions built from 
products of single-spin representations, and a corresponding
raster scan method in \cite{rundle2017} utilizes the 
probability to find the rotated system in each of the basis states 
of a Stern-Gerlach-type experiment.
In \cite{Schmied2011},
a similar tomographic method 
based on
filtered back projections (in analogy to planar inverse Radon 
transforms used e.g.\ in medical imaging)
is used 
in Stern-Gerlach-type experiments. 
However, complementary to these tomographic reconstruction schemes
for density matrices relying on Stern-Gerlach-type experiments,
our scheme is based on projections of operators onto rotated 
axial tensor operators or experimentally accessible expectation 
values of transformed axial tensor operators.
The measurement of spherical harmonics components of
electromagnetic near-field radiation using specifically designed loop 
antennas as sensors \cite{Vincent2009, Vincent2010, Breard2016}
is closely related to our interpretation of droplets as multipole potentials
(see Sec.~\ref{interpretation}).

Further tomographic approaches 
have been established in \cite{AT11,MJS12}
with applications to molecular systems. 
In those works, the quantum state given as 
the wave functions of an excited state
is obtained by decomposing the wave function in a series of basis functions 
and the expansion coefficients are acquired
by calculating a set of Fourier integrals 
from the detected signal. Also, the wave packet is 
reconstructed in \cite{HC04}
by computing the overlap of the state with well-defined reference states for different time intervals.

We want to also contrast our Wigner tomography approach
to the so-called spherical tensor analysis (STA) method
developed by Suter and Pearson \cite{SUTER1988} and 
van Beek et al. \cite{Levitt2005}. Both Wigner tomography and STA experiments have a 
similar structure. First, a preparation block (called an excitation sequence in STA) 
is used to prepare a density operator. In a second step, 
rotations around several axes are applied in a rotation block.
Finally, the density matrix is transformed 
(by a reconversion sequence in STA)
into a detectable basis and the signal is 
detected during an acquisition period.
Despite these similarities and the fact that both methods are based on 
characteristic properties of spherical tensor operators under rotations, 
the desired data and therefore also the details of the experiments 
differ considerably. 
The goal of STA is not to measure a density operator represented by 
spherical functions, but to decompose the detectable signal at a later time  
into individual signal components depending on 
occurring ranks $j$ and orders $m$
in the current density matrix.
In the standard form of STA \cite{SUTER1988,Levitt2005}, the detected signal 
of a density matrix is not characterized or decomposed
in terms of additional quantum numbers or labels $\ell$.
While a single reconversion sequence is used in STA, 
different pulse sequences are applied 
in the Wigner tomography in order to
transform operators to directly detectable ones 
(see \ Fig.~\ref{Tomo_Fig}(c), (d) and Table~\ref{tab:det_sequ}).
As a final difference, a rotation block in STA uses rotations for 
three Euler angles $\alpha$, $\beta$, and $\gamma$,
whereas in the Wigner tomography only two Euler
angles $\alpha$ and $\beta$ are necessary.

Lastly, our Wigner tomography can be seen as a stepping
stone along the path to identifying 
and characterizing operators in terms
of expansion coefficients for a suitable chosen basis.
One can consider various different bases such as 
simple matrix coefficients, a spherical tensor basis \cite{Garon15,SUTER1988}
(as the one used here), 
or a Cartesian product basis 
\cite{EBW87,Chuang447,PhysRevLett.80.3408, PhysRevA.67.062304, PhysRevA.69.052302}.
In this context of quantum state tomography, the shapes of 
spherical functions recovered in the Wigner tomography 
clearly contain highly redundant information,
but they also provide information about random or systematic errors of the tomography process itself.
One can also envision Wigner tomography as a component 
of a more general approach where
one would like
to optimize the number and location of samples for
achieving a desired fidelity 
and robustness against experimental errors \cite{Miranowicz2014,Rouve2006}, or
recover a physical density matrix and estimate experimental errors
(see, e.g., \cite{schwemmer15,knips15,faist16,silva17,steffens17,riofrio17,suess16} and references therein).

%-----------------------------------------------------------------------------------------------------
%---- aknowledgement
%-----------------------------------------------------------------------------------------------------

\begin{acknowledgments}
This work was supported in part by the Excellence Network of Bavaria (ENB) through ExQM. R.Z. 
and  S.J.G.  acknowledge  support  from  the  Deutsche Forschungsgemeinschaft (DFG) through
Grant No.\ Gl 203/7-2. We thank Raimund Marx for providing the samples. 
The experiments were performed at the Bavarian NMR Center at the Technical University of Munich.
\end{acknowledgments}

%-----------------------------------------------------------------------------------------------------
%---- appendix
%-----------------------------------------------------------------------------------------------------

\appendix

\section{Proof of Result~\ref{result1}\label{sec:proof_sf}}
We detail now the arguments leading to our reconstruction formula for a  
spherical function $g(\theta,\phi)$ as stated in Result~\ref{result1}.
This result relies on projections of 
rotated spherical harmonics.

We use the notation introduced in Sec.~\ref{drops_tomo}. The angles $\theta$ and $\phi$
indicate generic argument values of a spherical function $g(\theta,\phi)$, but the angles $\beta$ and $\alpha$ refer to
specific argument values. First, the right hand side 
$s_j\, \langle R_{\alpha\beta} Y_{j0}(\theta,\phi) | g(\theta,\phi)\rangle_{L^2}$
of Eq.~\eqref{tomoprimeaa} in Result~\ref{result1} is 
rewritten as 
$s_j\,
\langle R_{\alpha\beta} Y_{j0}(\theta,\phi) | g_j(\theta,\phi)\rangle_{L^2}$,
where the familiar orthogonality relation
$\langle Y_{jm}(\theta,\phi) |  Y_{j'm'}(\theta,\phi)\rangle_{L^2} = \delta_{jj'} \delta_{mm'}$
of spherical harmonics (see\  p.~68 of \cite{BL81}) implies the relation
$\langle g_j(\theta,\phi) |  g_{j'}(\theta,\phi)\rangle_{L^2} = \delta_{jj'}$ for the
rank-$j$ parts $ g_{j}(\theta,\phi)$ in the decomposition $g(\theta,\phi)=\sum_{j} g_{j}(\theta,\phi)$.
Second, 
one obtains that
$s_j\,
\langle R_{\alpha\beta} Y_{j0}(\theta,\phi) | g_j(\theta,\phi)\rangle_{L^2}
=s_j\,
\langle  Y_{j0}(\theta,\phi) |
R_{\alpha\beta}^{-1}
g_j(\theta,\phi)\rangle_{L^2}$ holds, which can be deduced from
the invariance
$\langle R_{\alpha\beta} Y_{jm}(\theta,\phi) | 
R_{\alpha\beta} Y_{j'm'}(\theta,\phi)\rangle_{L^2}
= \langle Y_{jm}(\theta,\phi) | 
Y_{j'm'}(\theta,\phi)\rangle_{L^2}
$
under rotations. The last relation
is easily verified using
the formula $R_{\alpha\beta} Y_{jm}(\theta,\phi)=
Y_{jm}(\theta{-}\beta,\phi{-}\alpha)$ and
a change of variables in the integral defining the scalar product
(see Sec.~\ref{drops_tomo}). Finally, 
$R_{\alpha\beta}^{-1}
g_j(\theta,\phi)$ is expanded into a linear combination 
$\sum_{m'=-j}^{j} c_{jm'}(\alpha,\beta) 
Y_{jm'}(\theta,\phi)$ 
of spherical harmonics
\footnote{Even though our arguments do not rely on explicitly knowing
the expansion coefficients $c_{jm'}(\alpha,\beta)$, it might be 
instructive to specify the expansion coefficients in terms
of the widely used Wigner-$D$ matrices $D_{m'm}^{j}(\alpha,\beta,\gamma)$ 
\cite{BL81,SR73}.
Let us also 
assume that the rank-$j$
parts
$g_j(\theta,\phi)$ 
are expanded into
$g_j(\theta,\phi)\allowbreak=\allowbreak\sum_{m=-j}^{j} \tilde{c}_{jm} Y_{jm}(\theta,\phi)$
using certain coefficients $\tilde{c}_{jm}$.
The theory of Wigner-$D$ matrices \cite{BL81,SR73}
implies that
$R_{\alpha\beta}^{-1}\, \allowbreak Y_{jm}(\theta,\phi) \allowbreak =\allowbreak
\sum_{m'=-j}^{j} \allowbreak
D_{m'm}^{j}(0,-\beta,-\alpha)\, \allowbreak Y_{jm'}(\theta,\phi)$.
Hence, we obtain the formula
$R_{\alpha\beta}^{-1}\, g_j(\theta,\phi) 
=\sum_{m=-j}^{j} \allowbreak \tilde{c}_{jm} \allowbreak \times \sum_{m'=-j}^{j} \allowbreak
D_{m'm}^{j}(0,-\beta,-\alpha)\, \allowbreak Y_{jm'}(\theta,\phi)$
which shows
that $c_{jm'}(\alpha,\beta) = \sum_{m=-j}^{j} \tilde{c}_{jm}
D_{m'm}^{j}(0,-\beta,-\alpha)$.}.
\nocite{BL81,SR73}
It follows that 
the right hand side of Eq.~\eqref{tomoprimeaa}
is given by
$s_j\,
\langle  Y_{j0}(\theta,\phi) |
\sum_{m'=-j}^{j} c_{jm'}(\alpha,\beta) 
Y_{jm'}(\theta,\phi)
\rangle_{L^2}=s_j\, c_{j0}(\alpha,\beta)$.

\begin{table}[b]
\caption{Sequences to prepare the density matrix $\rho$ from the thermal equilibrium state $\rho_{th}$, note 
$I_{abc}:=I_{1a}I_{2b}I_{3c}$. \label{tab:prep_sequ} }
\begin{tabular}{@{\hspace{0mm}} c @{\hspace{1.8mm}} c @{\hspace{1mm}} c @{\hspace{0mm}}} 
\\[-2mm]
\hline\hline
\\[-3mm]
No.\ of &
\\
spins & Sequence & $\rho$
\\[1mm]  \hline \\[-2mm]
%-------------------------------------------------------------------------------
1 & $[\frac{\pi}{2}]_{y}(I_1)$ & $I_x $ \\
& $[\frac{\pi}{2}]_{-x}(I_1)$ & $I_y $ \\
& identity operation (do nothing)
& $I_z $ \\
\noalign{\vskip 0.5em}
2 & ${\mathcal{P}}^{bil}_{x}$-$[\frac{\pi}{2}]_{y}(I_2)$ & $2I_{1x}I_{2x}$ \\
& ${\mathcal{P}}^{bil}_{y}$-$[\frac{\pi}{2}]_{-x}(I_2)$ & $2I_{1y}I_{2y}$ \\
& ${\mathcal{P}}^{bil}_{x}$-$[\frac{\pi}{2}]_{-y}(I_1)$ & $2I_{1z}I_{2z}$ \\
& ${\mathcal{P}}^{bil}_{x}$-$[\frac{\pi}{2}]_{-x}(I_2)$ & $2I_{1x}I_{2y}$ \\
& ${\mathcal{P}}^{bil}_{y}$-$[\frac{\pi}{2}]_{y}(I_2)$ & $2I_{1y}I_{2x}$ \\
&${\mathcal{P}}^{bil}_{x}$-$[\frac{\pi}{2}]_{y}(I_1)$-$[\frac{\pi}{2}]_{-y}(I_2)$ 
& $2I_{1z}I_{2x}$ \\
\noalign{\vskip 0.5em}
3 & ${\mathcal{P}}^{tril}_{x}$-$[\frac{\pi}{2}]_{y}(I_2)$-$[\frac{\pi}{2}]_{y}(I_3)$ 
& $4I_{xxx}$ \\
& ${\mathcal{P}}^{tril}_{x}$-$[\frac{\pi}{2}]_{-y}(I_1)$-$[\frac{\pi}{2}]_{-x}(\hspace{-0.2mm}I_1\hspace{-0.2mm})$-$[\hspace{-0.2mm}\frac{\pi}{2}\hspace{-0.2mm}]_{-x}(\hspace{-0.2mm}I_2\hspace{-0.2mm})$-$[\hspace{-0.2mm}\frac{\pi}{2}\hspace{-0.2mm}]_{-x}(\hspace{-0.2mm}I_3\hspace{-0.2mm})$ & $4I_{yyy}$ \\
& ${\mathcal{P}}^{tril}_{x}$-$[\frac{\pi}{2}]_{-x}(I_2)$-$[\frac{\pi}{2}]_{-x}(I_3)$ & $4I_{xyy}$ \\
& ${\mathcal{P}}^{tril}_{x}$-$[\frac{\pi}{2}]_{-y}(I_1)$-$[\frac{\pi}{2}]_{-x}(I_1)$-$[\frac{\pi}{2}]_{y}(I_2)$-$[\frac{\pi}{2}]_{-x}(I_3)$ 
& $4I_{yxy}$ \\
& ${\mathcal{P}}^{tril}_{x}$-$[\frac{\pi}{2}]_{-y}(I_1)$-$[\frac{\pi}{2}]_{-x}(I_1)$-$[\frac{\pi}{2}]_{-x}(I_2)$-$[\frac{\pi}{2}]_{y}(I_3)$ 
& $4I_{yyx}$ \\
& ${\mathcal{P}}^{tril}_{x}$-$[\frac{\pi}{2}]_{y}(I_2)$-$[\frac{\pi}{2}]_{-x}(I_3)$ & $4I_{xxy}$ \\
& ${\mathcal{P}}^{tril}_{x}$-$[\frac{\pi}{2}]_{-x}(I_2)$-$[\frac{\pi}{2}]_{y}(I_3)$ & $4I_{xyx}$ \\
& ${\mathcal{P}}^{tril}_{x}$-$[\frac{\pi}{2}]_{-y}(I_1)$-$[\frac{\pi}{2}]_{-x}(I_1)$-$[\frac{\pi}{2}]_{y}(I_2)$-$[\frac{\pi}{2}]_{y}(I_3)$ & $4I_{yxx}$ \\
& ${\mathcal{P}}^{tril}_{x}$-$[\frac{\pi}{2}]_{-x}(I_2)$ & $4I_{xyz}$
%-------------------------------------------------------------------------------
\\[1mm]  \hline \hline  \\[-2mm]
\end{tabular}
\end{table}

\begin{table}
\caption{Cartesian product operators $C_j^{(\ell,n)}$ with 
NMR-measurable operators $M_j^{(\ell,n)}$ used in the experiments, note $I_{abc}:=I_{1a}I_{2b}I_{3c}$. 
\label{tab:det_sequ} }
\begin{tabular}{@{\hspace{1mm}} l @{\hspace{7mm}} r @{\hspace{1mm}}}
\\[-2mm]
\hline\hline
\\[-2mm]
\begin{tabular}[t]{
@{\hspace{0mm}} c @{\hspace{2mm}} c @{\hspace{2mm}} c @{\hspace{3mm}}  c  @{\hspace{3mm}} c 
@{\hspace{0mm}} }
$\ell$ & $j$ & $n$ & $C_j^{(\ell,n)}$ &  $M_j^{(\ell,n)}$ 
\\[1mm]  \hline \\[-2mm]
%-------------------------------------------------------------------------------
$\{k\}$ & $1$ & $1$ & $I_{kz}$  & $I_{kx}$ \\
\noalign{\vskip 0.5em}
$\{kl\}$ & $0, 2$ & $1$ & $2I_{kx}I_{lx}$  & $2I_{kx}I_{lz}$ \\
&  & $2$ & $2I_{ky}I_{ly}$  & $2I_{ky}I_{lz}$ \\
&  & $3$ & $2I_{kz}I_{lz}$ &  $2I_{ky}I_{lz}$ \\
\noalign{\vskip 0.5em} 
& $1$ & $1$ & $2I_{kx}I_{ly}$ &  $2I_{kx}I_{lz}$ \\
&  & $2$ & $2I_{ky}I_{lx}$ &  $2I_{ky}I_{lz}$ \\
\noalign{\vskip 0.5em} 
$\tau_1$ & $1, 3$ & $1$ & $4I_{xxz}$ & $4I_{xzz}$ \\
& & $2$ & $4I_{xzx}$ & $4I_{xzz}$ \\
& & $3$ & $4I_{yyz}$ & $4I_{yzz}$ \\
& & $4$ & $4I_{yzy}$ & $4I_{yzz}$ \\
& & $5$ & $4I_{zxx}$ & $4I_{xzz}$ \\
& & $6$ & $4I_{zyy}$ & $4I_{xzz}$ \\
& & $7$ & $I_{zzz}$ & $4I_{xzz}$ \\
\noalign{\vskip 0.5em} 
$\tau_2$ & $1$ & $1$ & $4I_{xxz}$ & $4I_{xzz}$ \\
& & $2$ & $4I_{xzx}$ & $4I_{xzz}$ \\
& & $3$ & $4I_{yyz}$ & $4I_{yzz}$ \\
& & $4$ & $4I_{yzy}$ & $4I_{yzz}$ \\
& & $5$ & $4I_{zxx}$ & $4I_{xzz}$ \\
& & $6$ & $4I_{zyy}$ & $4I_{xzz}$ \\ 
\end{tabular}
&
\begin{tabular}[t]{
@{\hspace{0mm}} c @{\hspace{2mm}} c @{\hspace{2mm}} c @{\hspace{3mm}}  c  @{\hspace{3mm}} c 
@{\hspace{0mm}} }
$\ell$ & $j$ & $n$ & $C_j^{(\ell,n)}$ &  $M_j^{(\ell,n)}$ 
\\[1mm]  \hline \\[-2mm]
$\tau_2$ & $2$ & $1$ & $4I_{xzy}$ & $4I_{xzz}$ \\
& & $2$ & $4I_{yzx}$ & $4I_{yzz}$ \\
& & $3$ & $4I_{zxy}$ & $4I_{xzz}$ \\
& & $4$ & $4I_{zyx}$ & $4I_{xzz}$ \\
\noalign{\vskip 0.5em} 
$\tau_3$ & $1$ & $1$ & $4I_{xzx}$ & $4I_{xzz}$ \\
& & $2$ & $4I_{yzy}$ & $4I_{yzz}$ \\
& & $3$ & $4I_{zxx}$ & $4I_{xzz}$ \\
& & $4$ & $4I_{zyy}$ & $4I_{xzz}$ \\
\noalign{\vskip 0.5em} 
& $2$ & $1$ & $4I_{xyz}$ & $4I_{xzz}$ \\
& & $2$ & $4I_{xzy}$ & $4I_{xzz}$ \\
& & $3$ & $4I_{yxz}$ & $4I_{yzz}$ \\
& & $4$ & $4I_{yzx}$ & $4I_{yzz}$ \\
& & $5$ & $4I_{zxy}$ & $4I_{xzz}$ \\
& & $6$ & $4I_{zyx}$ & $4I_{xzz}$ \\
\noalign{\vskip 0.5em} 
$\tau_4$ & $0$ & $1$ &  $4I_{xyz}$ & $4I_{xzz}$ \\
& & $2$ & $4I_{xzy}$ & $4I_{xzz}$ \\
& & $3$ & $4I_{yxz}$ & $4I_{yzz}$ \\
& & $4$ & $4I_{yzx}$ & $4I_{yzz}$ \\
& & $5$ & $4I_{zxy}$ & $4I_{xzz}$ \\
& & $6$ & $4I_{zyx}$ & $4I_{xzz}$ \\[1mm]
\end{tabular}
\\[1mm]  \hline \hline  \\[-2mm]
\end{tabular}
\end{table}

The left hand side of Eq.~\eqref{tomoprimeaa}
is transformed into $g_j(\beta,\alpha)=R_{\alpha\beta}^{-1}
g_j(0,0) = \sum_{m'=-j}^{j} c_{jm'}(\alpha,\beta) Y_{jm'}(0,0)
= s_j\, c_{j0}(\alpha,\beta)$, where  the formula
$Y_{jm'}(0,\phi) = s_j \delta_{m'0}$ 
(see\ p.~16 of \cite{SR73}) has been applied. In summary, we have verified that both sides
of Eq.~\eqref{tomoprimeaa} agree, which completes the proof of Result~\ref{result1}.

\section{Proof of Result~\ref{result2}\label{proof_tens}}

In this appendix, we demonstrate the tomography formula 
as given in Result~\ref{result2} for an operator $A$ by 
applying the reconstruction formula of Result~\ref{result1}. The proof relies on mapping
$A$ to (a set of) spherical functions
$f^{(\ell)}(\theta,\phi)$ as detailed in Sec.~\ref{drops_sum}.

By substituting $g(\theta,\phi)$ with  $f^{(\ell)}(\theta,\phi)$ 
in Result~\ref{result1}
for each label $\ell$ separately, one obtains that
$f_j^{(\ell)}(\beta, \alpha)   \allowbreak = \allowbreak   s_j\,
\langle R_{\alpha\beta} Y_{j0}(\theta,\phi) | f^{(\ell)}(\theta,\phi)\rangle_{L^2}$.
Note that $\langle A^{(\ell)} | B^{(\ell)} \rangle
=\langle f_A^{(\ell)} | f_B^{(\ell)} \rangle_{L^2}$ for the 
spherical functions $f_A^{(\ell)}$ and $f_B^{(\ell)}$ corresponding to the 
operators $A$ and $B$, which can easily be  verified by expanding the 
arguments into tensor operators and spherical harmonics and
applying their orthonormality relations. Moreover,
the correspondence between operators and spherical functions
is covariant under rotations (see\ Proposition 1(d) in \cite{Garon15}), i.e.,
the operator $\mathcal{R}_{\alpha\beta} T_{j0}^{(\ell)}$ is mapped to
$R_{\alpha\beta} Y_{j0}(\theta,\phi)$ 
\footnote{
This relation can also be established using the theory of Wigner-$D$ matrices \cite{BL81,SR73}
where 
$D^{j}_{m'm}(\alpha,\beta,\gamma) 
= \protect{\langle} Y_{jm'}(\theta,\phi) | R_{\alpha\beta\gamma} 
Y_{jm}(\theta,\phi) \protect{\rangle_{L^2}}$ (see p.~41 and 276 of \cite{BL81})
and $D_{m'm}^{j}(\alpha,\beta,\gamma) = 
\protect{\langle} T_{jm'}^{(\ell)} | \mathcal{R}_{\alpha\beta\gamma} T_{jm}^{(\ell)} \protect{\rangle}$
(see p.~45 of \cite{BL81}).
Here, 
$\mathcal{R}_{\alpha\beta\gamma} C := R_{\alpha\beta\gamma} C { R}^{-1}_{\alpha\beta\gamma}$
with 
$R_{\alpha\beta\gamma}:=
e^{- i \alpha F_z} e^{- i \beta F_y} e^{- i \gamma F_z}$, 
cf.\ Sec.~\ref{drops_tomo}
where 
$\mathcal{R}_{\alpha\beta} = \mathcal{R}_{\alpha\beta 0}$.}.
The last two statements imply that $f_j^{(\ell)}(\beta, \alpha)   \allowbreak = \allowbreak   s_j\,
\langle \mathcal{R}_{\alpha\beta} T_{j0}^{(\ell)} | A^{(\ell)}\rangle
= \allowbreak   s_j\,
\langle \mathcal{R}_{\alpha\beta} T_{j0}^{(\ell)} | A\rangle
$, where the last step follows as $\langle A^{(\ell)} | A^{(\ell')} \rangle = 0$ if $\ell \neq \ell'$
(which is a consequence of the orthonormality of the tensor operators $T_{jm}^{(\ell)}$). This 
completes the proof of Eq.~\eqref{tomoprime} in Results~\ref{result2}, and 
Eq.~\eqref{exptomoprime} is then a direct consequence
due to the fact that $\mathcal{R}_{\alpha\beta} T_{j0}^{(\ell)}=
[\mathcal{R}_{\alpha\beta} T_{j0}^{(\ell)}]^\dagger$ is Hermitian.

\section{Derivation of Eq.~\eqref{tomozzz}}
\label{sec_drops_NMR_spec}
Here, we derive the formula of Eq.~\eqref{tomozzz}
given in Result~\ref{result3}
starting from Result~\ref{result2}. 
In standard NMR experiments, only the signatures of
Cartesian product operators \cite{EBW87} 
that contain a single transverse Cartesian operator $I_{ka}$ with $a\in \{ x, \ y\}$
(such as $I_{ka}$, $2I_{ka}I_{lz}$, and $4I_{ka}I_{lz} I_{mz}$)
can be measured 
\emph{directly}, and hence the expectation values of axial operators $T^{(\ell)}_{j0}$ are not {\it directly} 
accessible. 
Nevertheless, these expectation values can be measured {\it indirectly}
since the operators $T^{(\ell)}_{j0}=  \sum_n r^{(\ell,n)}_{j} C^{(\ell,n)}_{j}$
can always be expressed as real linear combinations of (Hermitian) standard 
Cartesian product operators $C^{(\ell,n)}_{j}$ \cite{EBW87}. Thus, 
the tomography formula of Eq.~\eqref{tomozzz} can be rewritten as 
$f^{(\ell)}_j(\beta, \alpha)=    s_j   \sum_n  r^{(\ell,n)}_{j}  \langle  {\cal R}_{\alpha \beta}  
C^{(\ell,n)}_{j} \rangle_\rho$. One obtains that
$f^{(\ell)}_j(\beta, \alpha)=s_j   \sum_n  r^{(\ell,n)}_{j}  {\rm Tr} \{    C^{(\ell,n)}_{j} {\tilde \rho}\}$ 
where ${\tilde \rho}:={ \mathfrak{R}}^\dagger_{\alpha \beta} \rho { \mathfrak{R}}_{\alpha \beta}$
by exploiting the action 
${\cal R}_{\alpha \beta}  C^{(\ell,n)}_{j} = { \mathfrak{R}}_{\alpha \beta}  C^{(\ell,n)}_{j} { \mathfrak{R}}^\dagger_{\alpha \beta}$
via the operator ${ \mathfrak{R}}_{\alpha \beta}:=\exp(- i \alpha \sum_{k=1}^n I_{kz}) \exp(- i \beta \sum_{k=1}^n I_{ky})$
and the fact that the trace is invariant under cyclic permutations.
The operators $C^{(\ell,n)}_{j}$ can be transformed into measurable operators 
$M^{(\ell,n)}_{j}= {U}^{(\ell,n)}_{j} C^{(\ell,n)}_{j} {U}^{(\ell,n)\dagger}_{j} $
with
unitary transformations
${U}^{(\ell,n)}_{j}$, which can be realized experimentally using radio-frequency pulses and 
coupling evolutions. Hence, the formula of Eq.~\eqref{tomozzz}
is given by $f^{(\ell)}_j(\beta, \alpha) =  s_j   
\sum_n r^{(\ell,n)}_{j}  {\rm Tr} \{  M^{(\ell,n)}_{j} {\tilde{\tilde \rho}}^{(\ell,j,n)}\}$, where  
immaterial  cyclic permutations of the trace have again been applied and where
${\tilde{\tilde \rho}}^{(\ell,j,n)}:={ U}^{(\ell,n)}_{j} {\tilde \rho} { U}^{(\ell,n)\dagger}_{j}$.
Finally, the definition of the expectation value yields 
$f^{(\ell)}_j(\beta, \alpha)= s_j \sum_n r^{(\ell,n)}_{j} \langle M^{(\ell,n)}_{j} \rangle_{\tilde{\tilde \rho}^{(\ell,j,n)}}$.

\section{Preparation and detection sequences \label{prep_detec}}
We detail the explicit form of the preparation and detection sequences
used in the experiments in order to demonstrate our Wigner tomography.
We denote a pulse with flip angle $\beta$ and phase $\alpha$ that is 
applied to spin $k$ by $[\beta]_\alpha(I_k)$. Similarly,
$[\beta]_\alpha(I_k, I_l)$
specifies two pulses both of flip angle $\beta$ and phase $\alpha$ that are 
simultaneously applied to spins $k$ and $l$.
We also use the notation 
${\mathcal{P}}^{bil}_{x,y}=[\frac{\pi}{2}]_{y}(I_2)$-$G$-$[\frac{\pi}{2}]_{x,y}(I_1)$-$ t_a$-
$[\pi]_{y,x}(I_1, I_2)$-$ t_a$  which represents a pulse sequence
which is read from left to right. Here, $G$ represents a pulsed magnetic field gradient 
that dephases all present transverse spin operators and $t_a$ refers to a
time delay of length $1/(4J_{12})$, where  $J_{12}$ is the coupling
constant between the first and second spin. The pulse sequences
${\mathcal{P}}^{bil}_{x}$ and ${\mathcal{P}}^{bil}_{y}$
create from $I_z$ the bilinear product operators $2 I_{1x}I_{2z}$ and $2 I_{1y}I_{2z}$, respectively.
Moreover, the trilinear product operator  $4I_{xzz}:=4I_{1x}I_{2z}I_{3z}$ is obtained from $I_z$
by applying the pulse sequence 
${\mathcal{P}}^{tril}_x=
[\frac{\pi}{2}]_{y}(I_2, I_3)$-$G$-$[\frac{\pi}{2}]_{y}(I_1)$-$t_b$-$ [\pi]_y(I_1, I_3)$-$t_c$-$[\pi]_y(I_2)$-$ t_d$ 
where the time delays are
$t_b=1/(4J_{13})$, $t_c=1/(4J_{13})-1/(4J_{12})$, and $t_d=1/(4 J_{12})$.
Using these notations, the preparation sequences are given in 
Table~\ref{tab:prep_sequ}.

For the detection, the Cartesian product operators $C_j^{(\ell,n)}$ 
have to be rotated into 
NMR-measurable operators $M_j^{(\ell,n)}$. The relevant pairs
of operators $C_j^{(\ell,n)}$ and $M_j^{(\ell,n)}$ are provided in
Table~\ref{tab:det_sequ}. The rotation pulse sequences 
are easily inferred, e.g., 
one uses the pulse $[\pi/2]_{y}(I_k)$
in order to rotate $I_{kz}$ into $I_{kx}$. Similarly,
$[\pi/2]_{-y}(I_k)$, $[\pi/2]_{x}(I_k)$, and $[\pi/2]_{-x}(I_k)$ rotate
respectively
$I_{kx}$, $I_{ky}$, and $I_{kz}$ into $I_{kz}$, $I_{kz}$, and $I_{ky}$.
For example, $4I_{zxy}$ is rotated into $4I_{xzz}$
using the pulse sequence $[\frac{\pi}{2}]_{x}(I_3)$-$[\frac{\pi}{2}]_{-y}(I_2)$-$[\frac{\pi}{2}]_{y}(I_1)$.

%-----------------------------------------------------------------------------------------------------%
%---- reference (bibtex)
%\bibliography{big}

\end{document}